\documentclass[useAMS,usenatbib]{mnras}
\usepackage{amsmath}
\usepackage{amssymb}
\usepackage{graphicx}
\usepackage{multirow}
\usepackage{color}

\title[X-ray reverberation using an extended corona model]{Investigating the X-ray time-lags in PG~1244+026 using an extended corona model}

\author[P. Chainakun and A. J. Young]{P. Chainakun\thanks{E-mail:
phxpc@bristol.ac.uk} and A. J. Young \\H. H. Wills Physics Laboratory, Tyndall Avenue, Bristol BS8 1TL}
 
\begin{document}

\date{}

\pagerange{\pageref{firstpage}--\pageref{lastpage}} \pubyear{2015}

\maketitle

\label{firstpage}

\begin{abstract}
We present an extended corona model based on ray-tracing simulations to investigate X-ray time lags in Active Galactic Nuclei (AGN). This model consists of two axial point sources illuminating an accretion disc that produce the reverberation lags. These lags are due to the time delays between the directly observed and reflection photons and are associated with the light-travel time between the source and the disc, so they allow us to probe the disc-corona geometry. We assume the variations of two X-ray sources are triggered by the same primary variations, but allow the two sources to respond in different ways (i.e. having different source responses). The variations of each source induce a delayed accretion disc response and the total lags consist of a combination of both source and disc responses. We show that the extended corona model can reproduce both the low-frequency hard and high-frequency soft (reverberation) lags. Fitting the model to the timing data of PG~1244+026 reveals the hard and soft X-ray sources at $\sim6r_{\text{g}}$ and $\sim11r_{\text{g}}$, respectively. The upper source produces small amounts of reflection and can be interpreted as a relativistic jet, or outflowing blob, whose emission is beamed away from the disc. This explains the observed lag-energy in which there is no soft lag at energies $<1$~keV as they are diluted by the soft continuum of the upper source. Finally, our models suggest that the fluctuations propagating between the two sources of PG~1244+026 are possible but only at near the speed of light.
\end{abstract}


\section{Introduction}

X-rays from accretion onto a super-massive black hole in Active Galactic Nuclei (AGN) provide a unique probe of the innermost region close to the central engine. The X-rays are produced in a corona via inverse Compton scattering of the optical and ultra-violet disc photons which can be observed either directly as the X-ray continuum, or back-scattered off the disc as the reflection component. The location of the corona (e.g. X-ray source height) produces the time delays between the direct and reflection components due to their different light-travel distances to an observer. Thus delays of the reflection dominated bands (e.g. the soft excess, Fe K and Compton hump bands) with respect to the continuum dominated bands are expected. These delays, or reverberation lags, were first observed in 1H0707-495 \citep{Fabian2009} and have been commonly used to determine the location of the X-ray source \citep[see][for a review]{Uttley2014}. 

The realistic reverberation lags can be modelled numerically using ray tracing techniques \citep{Karas1992, Fanton1997, Reynolds1999, Ruszkowski2000, Chainakun2012} by tracking the photon trajectories along Kerr geodesics and computing the arrival time of direct and reflection photons. \cite{Wilkins2013} employed the impulse response function to model the frequency-dependent time lags for different source geometries. \cite{Emmanoulopoulos2014} systematically fitted the lag-frequency spectra to the observational data under the lamp-post assumption. \cite{Cackett2014} investigated further how both frequency- and energy-dependent time lags in the Fe K band change with a wide range of key parameters such as the source height, inclination and black hole mass. Moreover, simultaneous fitting of the mean and lag spectra has been performed by \cite{Chainakun2015} and \cite{Chainakun2016} including the full effects of dilution and ionization gradients on the disc. \cite{Chainakun2016} also show that the lamp-post assumption can reproduce the important features of the observational lag-energy spectra in which the soft excess and the Fe K bands lag behind the continuum dominated band. More importantly, the dips of the lags at $\sim 3$ keV and $\sim 7$--10 keV are found to be naturally explained if the disc is highly ionized at the centre and is colder further out. \cite{Epitropakis2016b} adopted a theoretical model taking into account the full dilution effects, similar to \cite{Chainakun2015,Chainakun2016}, and fit the Fe K lags that have minimal bias, estimated using the new technique presented in \cite{Epitropakis2016}. They found that only the source height can be well constrained. Of all seven AGN they investigated, the X-ray source was found to be at a small distance ($\lesssim 10r_{\rm g}$) above the black hole.

However, AGN also exhibit low frequency hard lags that cannot easily be explained by X-ray reverberation. These hard lags occur on longer timescales than the timescales of the inner-disc reflection and are thought to be caused by the propagation of accretion rate fluctuations thorugh the disc \citep[e.g.][]{Arevalo2006}. If the X-ray emission region is harder towards the centre, the fluctuations propagating inwards will modulate the softer region first producing the hard lags. Recently, \cite{Wilkins2016} investigated time lags from an extended corona where the X-ray reverberation is driven by the causal propagation through the extended corona. Their model can qualitatively explain the observed features of time lags if the propagation is on the viscous timescales of the accretion disc. They also found the 3~keV dip is prominent when there is propagation up through the central collimated corona. This supports the hypothesis that the total time lags in AGN are produced by two different mechanisms: reverberation and propagating-fluctuation.     

Fitting time lags using a complex geometry such as an extended corona is very challenging and computationally intensive. In this paper we develop a self-consistent model using two X-ray point sources located on the rotation axis of the black hole, as an approximation of a vertically-extended source. Two X-ray sources are allowed to have different variability depending on how they respond to a primary, extrinsic variation. Although we do not model the propagating fluctuations, we employ a phenomenological function of the source responses that are expected if the sources are reacting to those propagating fluctuations. The variability of the X-ray continuum depends on the source response while the associated X-ray reflection depends also on the disc response. Therefore the model has to predict the time lags taking into account both continuum sources and the disc responses.

Ultimately, we aim to fit this ``two-blobs'' model to the timing data of the narrow-line Seyfert 1 galaxy PG~1244+026 $(z = 0.0482)$ observed with the \emph{XMM-Newton} satellite \citep{Jansen2001} on 2011 December 25 (ObsID 0675320101). The spectral analysis was carried out by \citet{Jin2013} who reported an additional, uncorrelated soft X-ray component. They found that spectral fitting alone cannot break degeneracies between the true separate soft-excess and relativistic reflection components. Therefore timing analysis is essential for this AGN. The light curve ($\sim 120$~ks) and data reduction here are similar to those have been analysed by \cite{Kara2014}. The lag-energy spectrum at the frequency range of $(0.9-3.6)\times10^{-4}$~Hz shows a strong Fe~K lag without the soft-excess lags. Although it may be that this frequency range covers the range whose positive hard and negative soft lags are taking place so that the propagation lags are switching to reverberation lags, these properties cannot be trivially explained by the standard lamp-post model. \cite{Kara2014} suggested that the soft-excess lags are diluted by the X-ray emission from the relativistic jet which is not correlated with the disc reflection. \cite{Alston2014} suggested that the separate soft-excess component should have its own lags that contribute to the total time lags. They found that taking into account the lags contributed by the blackbody emission helps improve the fits in the soft bands. It is interesting to see if our two-blobs model can provide more insights into the complex source geometry of PG~1244+026.

In the next Section we present essential equations and theoretical concepts relating to the two-blobs model. Our assumptions and methods for calculating reverberation lags are described in Section 2.1. In Section 2.2, we investigate more complex scenario in which two X-ray sources vary differently with respect to the primary variation and show that under these circumstances both low-frequency hard and high-frequency soft lags can be produced. To fit the lag-frequency and lag-energy spectra of PG~1244+026, we need to simplify the model so that it is efficient and not too time consuming to evaluate. The fitting procedure together with the fitting results are presented in Section 3. We discuss the results towards the proposed geometry of PG~1244+026 in Section 4. Our conclusions are presented in Section 5. 

\section{Two-blobs model}

\subsection{Modelling reverberation lags}

\begin{figure}
    \centering
    \includegraphics*[width=85mm]{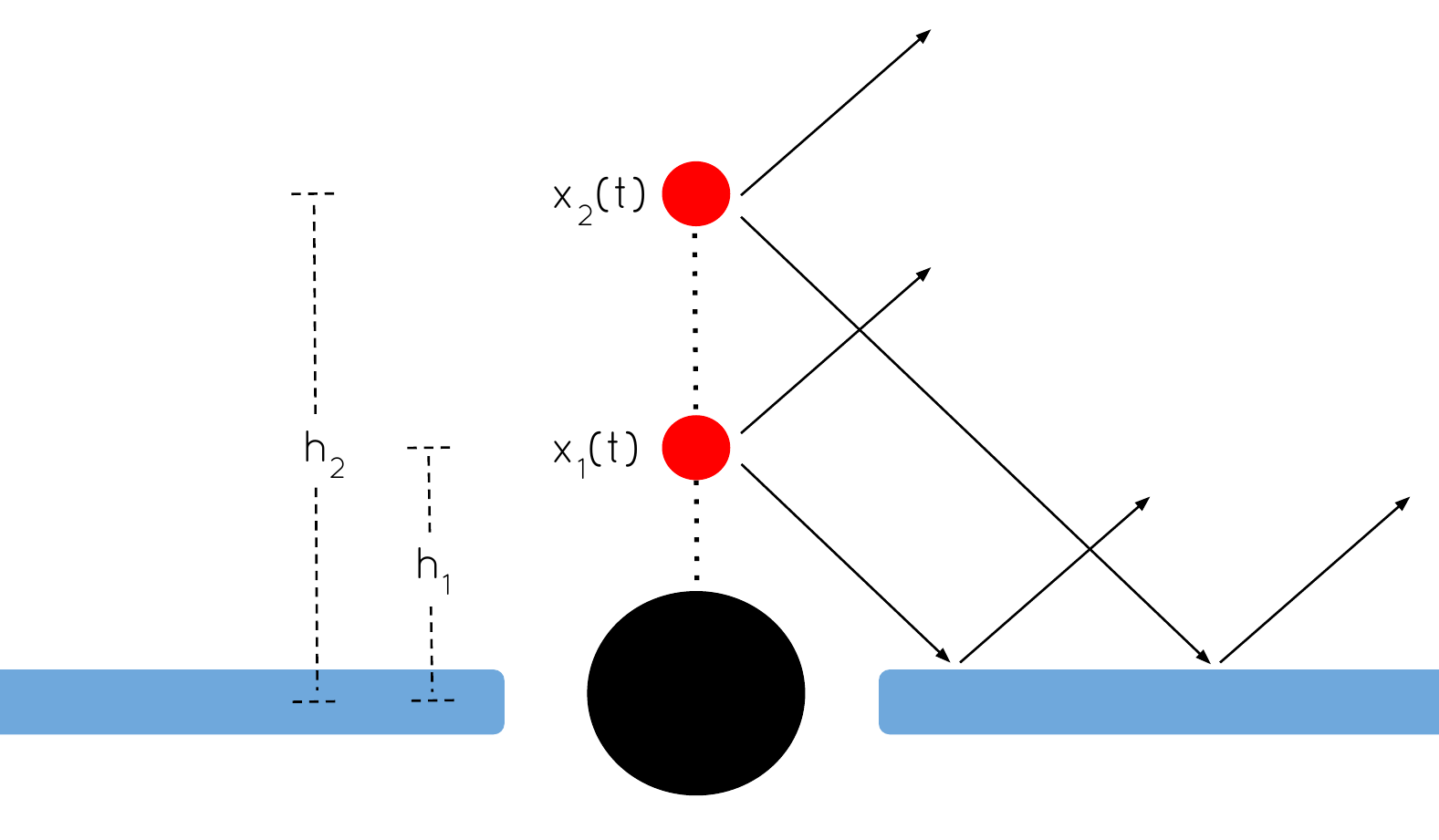}
    \caption{Sketch of the ``two-blobs'' model where two X-ray sources are located on the rotation axis of the black hole. We define $x_{i}(t)$ as the time dependent amplitude of the X-ray sources where the subscripts $i=1$ and 2 refer to the lower and upper sources, respectively. The source heights are $h_1$ and $h_2$.} 
    \label{pma}
\end{figure}

For all our model calculations we measure distance and time in gravitational units which are $r_{\text{g}} = GM/c^2$ and $t_{\text{g}} = GM/c^3$, respectively, where $G$ is the gravitational constant, $c$ is the speed of light and $M$ is the black hole mass. We assume the central black hole rotates with physically maximum spin, $a=0.998$, although the calculations can easily be performed for arbitrary values of $a$. The disc is a standard geometrically thin, optically thick disc \citep{Shakura1973} extending from the radius of the innermost stable circular orbit, or radius of marginal stability $r_{ms}$, to $400r_{\text{g}}$ and is illuminated by two X-ray point-sources. The lower and higher sources are located on the symmetry axis at heights $h_{1}$ and $h_{2}$ whose amplitudes as a function of time are given by $x_{1}(t)$ and $x_{2}(t)$, respectively. A sketch of the model geometry is shown in Fig.~\ref{pma}. 

The ray tracing technique is used to compute the impulse disc-response function \citep{Wilkins2013, Cackett2014, Emmanoulopoulos2014, Chainakun2015, Chainakun2016,Wilkins2016,Epitropakis2016b} or count rates of the reflection photons as a function of time after the direct continuum is detected. We first assume two delta-function flares as two sources of X-rays and trace emitted photons between the sources, the disc and the observer along Kerr geodesics. The redshifts and full-relativistic effects are included \citep[e.g.][]{Cunningham1975}. The disc has a radial-density profile in the form of a power-law, $n(r) \propto r^{-p}$ so that the ionization parameter is $\xi(r,\phi) = 4 \pi F_t(r,\phi) / n(r)$ where $F_t(r,\phi)$ is the total incident flux, due to both X-ray sources, per unit area of the $(r,\phi)$ disc segment, in the disc frame. We use $\xi_{ms}$ and $p$ as the model parameters that determine the ionization gradient of the disc. The first parameter, $\xi_{ms}$, is the averaged ionization state at $r_{ms}$ and how the disc becomes less ionized further out is determined by the disc density index, $p$. The X-ray reprocessing is modelled using {\sc reflionx} \citep{George1991, Ross1999, Ross2005} which requires the iron abundance, $A$, as an additional model parameter. Note that we can in principle use other reflection models, such as {\sc xillver} \citep[e.g.][]{Garcia2014}, which we intend to do in the future. Frequency-dependent time lags are usually measured between the reflection-dominated and continuum-dominated bands, and are diluted due to the contamination of the cross components (i.e. continuum contributing to the reflection band, and vice versa). The observed time lags then are always smaller than the intrinsic time lags because of the dilution effects \citep[see, e.g.,][for a discussion about dilution effects]{Uttley2014, Chainakun2015}.

Let us begin with the simplest case when both X-ray sources simultaneously and coherently respond to the primary variations $x_{0}(t)$. The spectrum is split up into energy bands $E_{j}$. The source variability is then
\begin{equation}
    x_{i}(E_{j},t) \propto F_{i}(E_{j})x_{0}(t)
    \label{eq:source_var}
\end{equation}
where the subscripts $i=1$ and 2 refer to the parameters of the lower and upper X-ray source, respectively. The flux across all energy bands is weighted by 
\begin{equation}
    F_{i}(E_{j}) \propto 
    \int_{E_{j,\rm low}}^{E_{j,\rm high}} {E_{j}^\prime}^{-\Gamma_i} {\rm d} E_{j}^\prime
    \label{eq:e_band}
\end{equation}
where $\Gamma_{i}$ is the photon index of the X-ray continuum. The limits are the energy range covered by the band $E_{j}$ and thus $x_{i}(E_{j},t)$ is proportional to, or is an estimate of, the direct X-ray variability in that band. This normalization is also employed by \cite{Emmanoulopoulos2014, Chainakun2015, Chainakun2016}. The variability of an echo from the disc is calculated via a convolution term,
\begin{equation}
    \begin{split}
        v_{i}(E_{j},t) & = x_{i}(E_{j^{*}},t') \otimes \psi_{i}(E_{j},t') \\
        & = \int_0^t x_{i}(E_{j^{*}},t^\prime) \psi_{i}(E_{j},t-t^\prime)dt^\prime
        \label{eq:response_lc}
    \end{split}
\end{equation}
where $\psi_{i}(E_{j},t')$ is the impulse disc-response function at time $t^\prime$ after the flare, obtained via the ray-tracing simulations of photons of all X-ray energies. The * indicates a mapping of the energies in the driving signal to the energies of response signal as the relativistic effects would also affect the distribution in energy. Since we consider energies in the driving signal that vary coherently and simultaneously (equation~\ref{eq:source_var}), the shifting in energies should not have a significant effect on the variability distribution in those bands and hence equation~\ref{eq:response_lc} without * would still be a good approximation. Nevertheless, we perform the full relativistic ray-tracing simulations so that this and other effects due to general relativity are taken into account and are implemented in the response function.

Each X-ray source is assumed to have its own disc-response function calculated with the same disc properties (i.e. $\xi_{ms}$, $p$ and $A$ are the same for two X-ray sources). The offset time zero of the disc-response functions is set to be the earliest time of which the direct photons are detected. We also assume the X-ray reprocessing in the disc occurs instantaneously so that the reverberation lags mainly depend on the source geometry and the amount of dilution. The observed X-ray variability can be written in a general form of 
\begin{equation}
    a(E_{j},t) = \sum\limits_{i=1}^{2} B_{i} \Big{(} x_{i}(E_{j^{*}},t)+R_{i}(E_{j})v_{i}(E_{j},t) \Big{)}
    \label{eq:obs_var}
\end{equation} 
where $B_{i}$ is the brightness parameter measured in the observer's frame and $R_{i}$ is the reflected response fraction. In our model, $B_{i}$ is applied not just to the continuum but also to the reflection (i.e. when the continuum flux increases with $B_{i}$, the reflection flux increases with the same factor as it is the continuum convolved with the disc response). However, if $B_{1}=B_{2}$ the lags will be the same regardless of what their values are. We then define the brightness ratio $B=B_{2}/B_{1}$ and always fix $B_{1}=1$ such that $B<1$ or $B>1$ is the case when the $2^{\text{nd}}$ X-ray source is seen to be fainter or brighter than the $1^{\text{st}}$ source, respectively. The emitted photons from the source closer to the black hole are more focused towards the centre due to the light bending effects \citep{Miniutti2004} and, as a result, a relatively low continuum flux is observed. We can estimate the level of continuum flux by counting the number of direct photons that escape to the infinity. The continuum flux increases more significantly compared to the reflection flux for a higher source height. We then expect $R_{2}<R_{1}$. Even though $B>1$ is also expected, as it is related to the continuum flux, but we are dealing with two X-ray sources whose intrinsic luminosity may not be the same. Low intrinsic luminosity of the upper source with respect to the lower source can lead to $B\ll 1$. Therefore $B$ is allowed to be a free parameter in our model. The equation~\ref{eq:obs_var} can also be generalised in a straightforward way to ``$n$'' sources, not just the two we are computing.

Furthermore, we define the parameter $R_{i}$ as the Reflected Response Fraction (RRF) which is (reflection flux)/(continuum flux) measured in the 5--7 keV band. There is no prefered energy band for $R_{i}$ because we are interested only in the relative fraction across all bands, $R_{i}(E_{j})$ \citep[see][for a discussion about RRF]{Chainakun2016}. The $R_{i}(E_{j})$ tells us the ratio (reflection flux)/(continuum flux) in the energy band $E_{j}$ that plays a role in diluting the lags of that band. $R_i(E_j)$ only needs to be specified in a single energy band because the values at all other energies can then be calculted self-consistently. The dilution effects are taken into account by normalizing the variability of the echo from the disc with $R_{i}(E_{j})$, whose value is known once the model parameter $R_{i}$ is selected. More importantly, since we use the entire energy spectrum of the X-ray continuum and reflection, the full dilution effects are self-consistently included in the model. This approach has recently been adopted by \cite{Chainakun2015, Chainakun2016,Wilkins2016,Epitropakis2016b}. 

\begin{figure}
    \centering
    \includegraphics*[width=70mm]{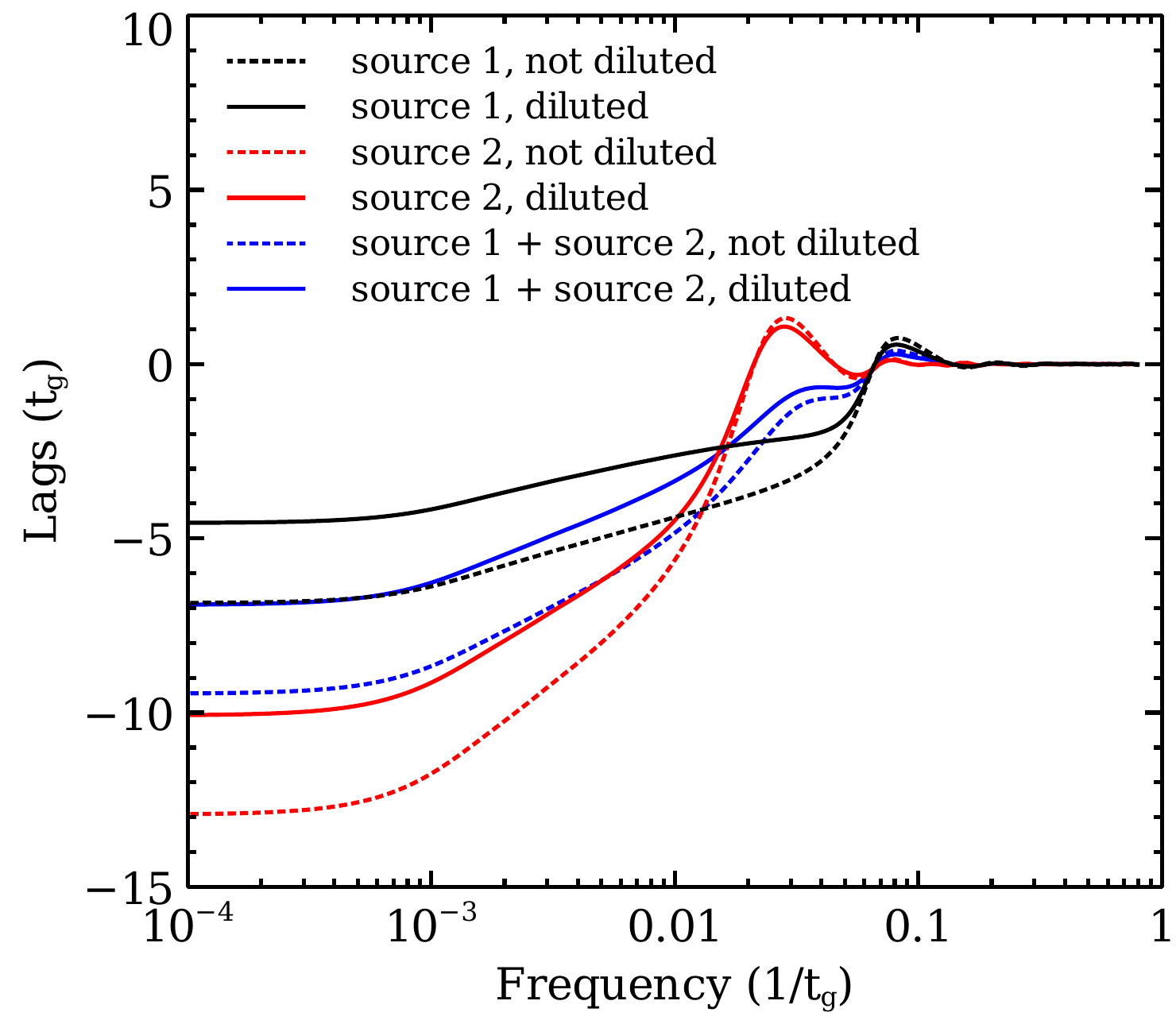}
    \caption{Frequency-dependent time lags between 0.3--1 keV and 2--4 keV bands comparing between the lamp-post and two-blobs solutions. The $1^{\text{st}}$ and $2^{\text{nd}}$ sources are located at $2r_{\text{g}}$ and $8r_{\text{g}}$, respectively. The black, red and blue solid lines represent the lags due to the $1^{\text{st}}$, the $2^{\text{nd}}$ and both X-ray sources, respectively. The corresponding time-lags without dilution are also presented in dashed lines.} 
    \label{p1}
\end{figure}

Time lags are calculated in the standard way using the argument of the cross-spectrum from the Fourier transforms of the light-curve pairs \citep{Nowak1999}. The negative and positive signs of time lags indicate the soft band lagging behind and leading the hard band, respectively. Fig.~\ref{p1} illustrates how the lags change subject to the existence of the $2^{\text{nd}}$ X-ray source. In this example we calculate the lags between the 0.3--1 vs. 2--4~keV bands (i.e. Fe L lags). Two X-ray sources are located at $h_{1}=2r_{\text{g}}$ and $h_{2}=8r_{\text{g}}$. The inclination $\theta=45^{\circ}$. The disc is assumed to have an iron abundance $A=1$ and the ionization at the innermost part $\xi_\text{ms} = 10^3 \text{ erg cm s}^{-1}$ which is colder further out with $p=0$. We set $\Gamma_{1}=\Gamma_{2}=2.0$, $R_{1}=1.0$, $R_{2}=0.5$ and $B=1$ (i.e. both X-ray sources have the same photon index and observed luminosity, but the reflection flux from the upper source is about a half of that from the lower source). The observed lags including dilution are always smaller than the intrinsic time lags which is important in measuring the true distance of the X-ray source. Dilution effects scale down the lags so the phase-wrapping frequency at which zero lag occurs does not change. The lags from two X-ray blobs are in between the lags produced separately by each blob. Although the results are not shown here, we find that if one of the X-ray sources is very bright or very faint, the two-blobs solution will converge to the lamp-post cases in which the disc is illuminated by only the single brightest source.

\subsection{Different source-responses}

Now we investigate the cases when the variations of two X-ray sources are triggered by the same primary variations, $x_{0}(t)$, but they respond in different ways. The impulse response of the $i^{\text{th}}$ source (aka, source response) is determined by the function $\Psi_{i}(t)$ that enforces the source variability,
\begin{equation}
    x_{i}(E_{j},t) \propto F_{i}(E_{j})\:x_{0}(t) \otimes \Psi_{i}(t).
    \label{eq:source_impulse_var}
\end{equation}
where $F_i(E_j)$ is the continuum flux in energy band $E_{j}$, defined in equation~\ref{eq:e_band}. The variation of each source then produces a delayed accretion disc response, $\psi_{i}(E_{j},t)$, which is obtained by the ray-tracing simulations. The observed X-ray variability in a specific energy band is then calculated using equation~\ref{eq:obs_var}.

\begin{figure}
    \centerline{
        \includegraphics*[width=0.45\textwidth]{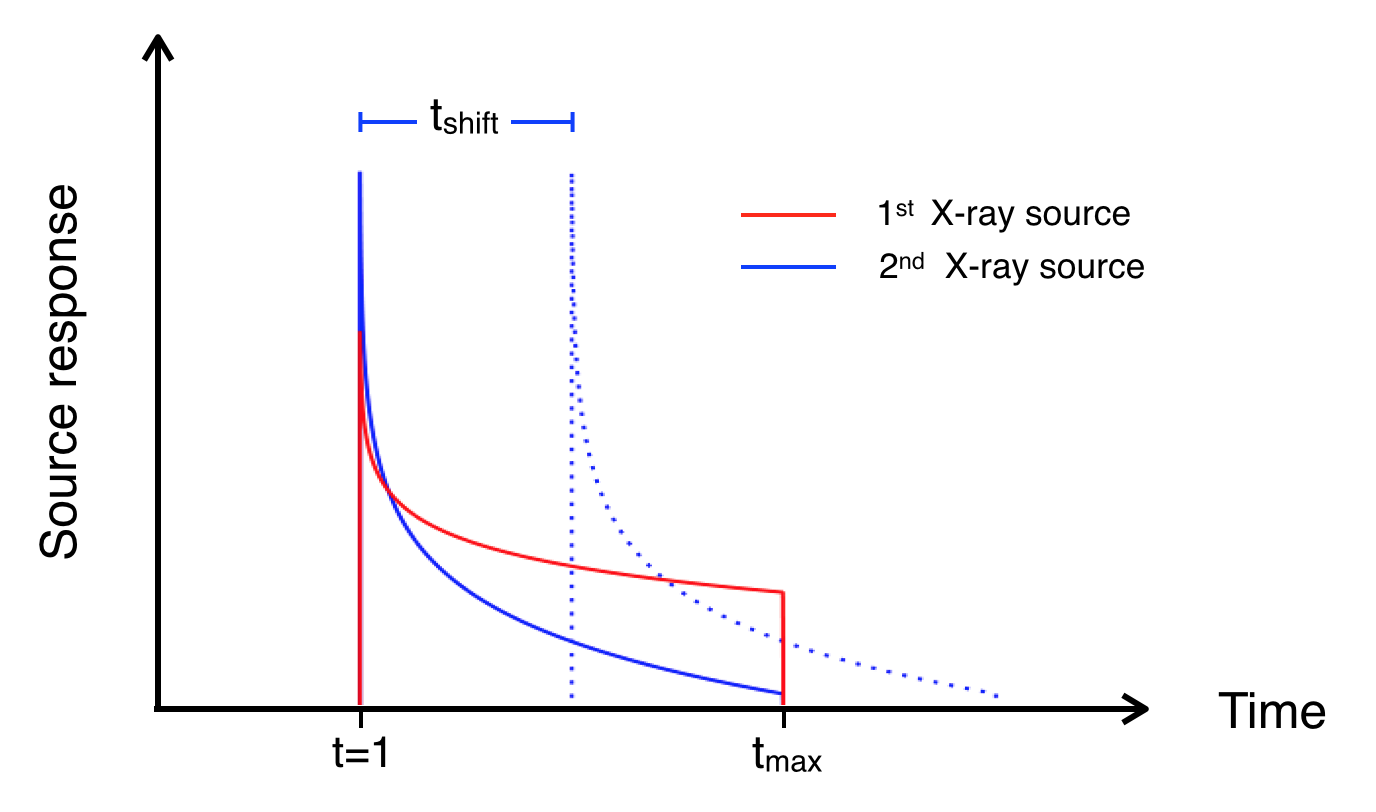}
    }
    \centerline{
        \includegraphics*[width=0.45\textwidth]{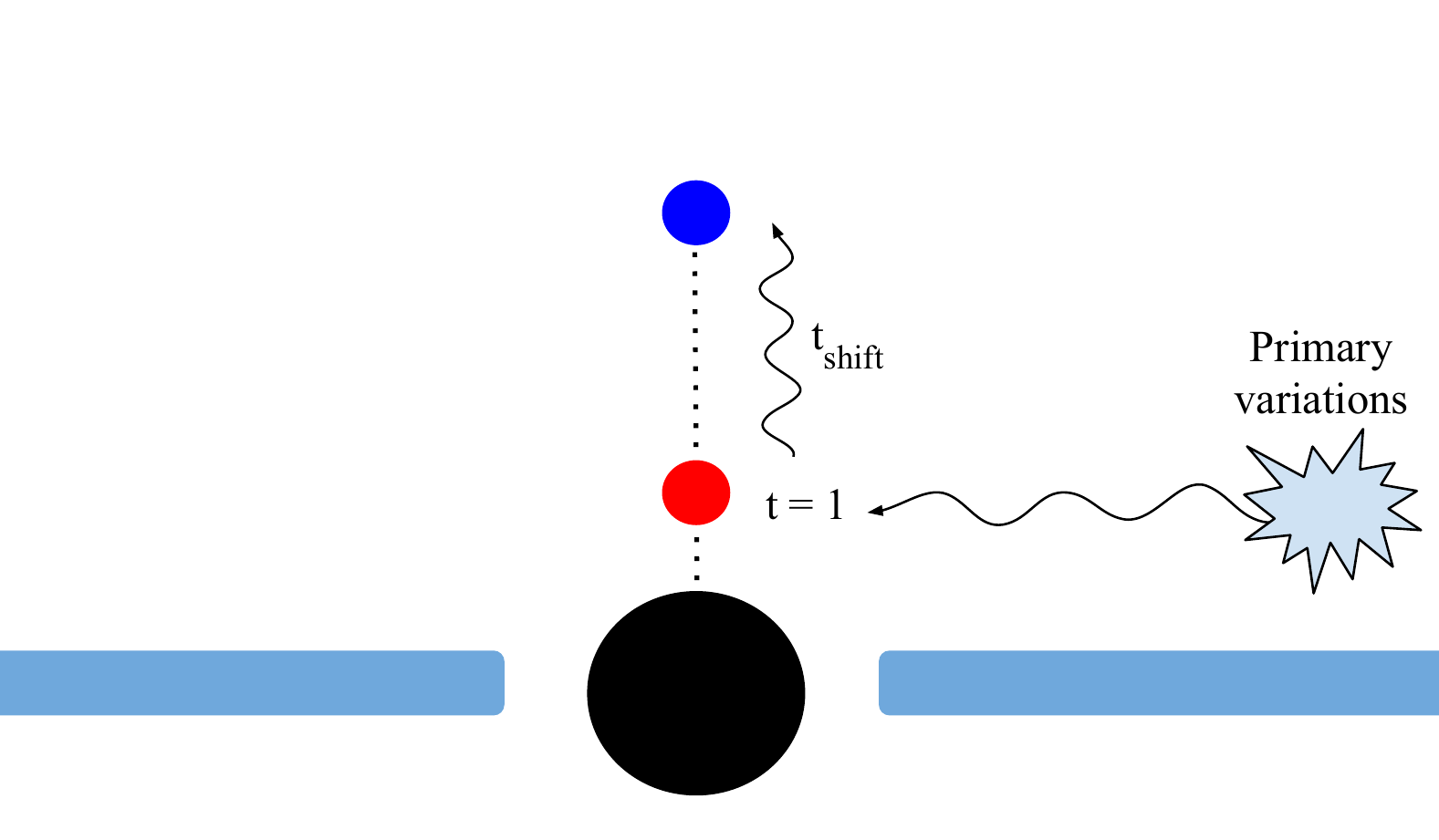}
    }
    \caption{Top panel: examples of the impulse source-response functions modelled using equation~\ref{eq:source_response}. The maximum responses are at $t=1$ and decrease with time towards the cut-off at $t_{\text{max}}$. The responses are shaped by the parameter $q$. We employ the parameter $t_{\text{shift}}$ to systematically delay one source response with respect to the other one. Note that even we set the initial response-time at $t=1$, there is no preferable choice because only relative difference between source responses is relevant. Bottom panel: one possible scenario consistent with our source-response assumption. See text for more details.}
    \label{sr}
\end{figure}

We aim to produce the low-frequency hard lags together with the high-frequency soft lags observed in many AGN. The low-frequency hard lags can be explained by mass accretion rate fluctuations which are propagating inwards \citep[e.g.][]{Lyubarskii1997,Kotov2001,Arevalo2006} where the X-ray sources are harder toward the centre. The fluctuations modulate the soft X-ray sources first and, as a result, hard lags occur. In other words the soft band flux dominates at first before the hard band flux gradually takes over. We then assume the propagating fluctuations affect the central X-ray sources in the way that one of the source responds more at early times before the other source responds more at later times. The simplest way to model the impulse source response is using a cut-off power-law:
 \begin{equation}
    \Psi_{i}(t) \propto t^{-q_{i}}\text{exp}(-t/t_{\text{max}}),
    \label{eq:source_response}
\end{equation}
where $t=1$ and $t_{\text{max}}$ are the beginning and the end of the response with the maximum and minimum responses, respectively. Note that there is no preferable choice for the initial response-time (which is set at $t=1$) because we are interested only in the relative difference between source responses. The parameter $q_{i}$ determines how the function decays over time. The area under the profile is normalized to 1. Examples of the modelled source responses are illustrated in Fig.~\ref{sr} (top panel). We also employ the parameter $t_{\text{shift}}$ to delay the response of the second X-ray source that reacts systematically slower to the primary variations than the first. Delaying one with $t_{\text{shift}}$ is similar to shifting the other source ahead in time by the same interval.    

Fig.~\ref{sr} (bottom panel) represents a physical scenario consistent with our source parameters. Primary variations are possibly the result of perturbation from the magneto-rotational instability \citep[e.g.][]{Balbus1991}. These produce fluctuations on the mass accretion rate which are propagated inwards \citep[e.g.][]{Arevalo2006} and first seen by the lower X-ray source at time $t=1$. The fluctuations then are propagated upwards from the lower to the upper source, taking time $t_{\text{shift}}$. The reverse situation in which downwards propagation occurs can also be modelled if required.

The full model has 16 parameters in total: the $1^{\text{st}}$ and $2^{\text{nd}}$ source height ($h_{1}$ and $h_{2}$), their photon indices ($\Gamma_{1}$, $\Gamma_{2}$), RRF ($R_{1}$, $R_{2}$), brightness ratio ($B$), inclination ($\theta$), iron abundance ($A$), ionization state at the $r_\text{ms}$ ($\xi_\text{ms}$), disc density index ($p$), black hole mass ($M$), source-response indices ($q_{1}$, $q_{2}$), maximum source-response time ($t_{\text{max}}$) and initial time shift ($t_{\text{shift}}$). To simplify the problem, while the disc response is a function of energy and time, the source response is energy-independent. The source variability, however, as shown in equation~\ref{eq:source_impulse_var} is still energy-dependent due to the factor $F_{i}(E)$. Note that there is no preferable energy band to be selected for the parameters $R_{1}$ and $R_{2}$. We use the 5--7~keV band and once the values of $R_{1}$ and $R_{2}$ are assumed they will reveal the corresponding RRF across all energy bands.

\begin{figure}
    \centering  
    \includegraphics*[width=70mm]{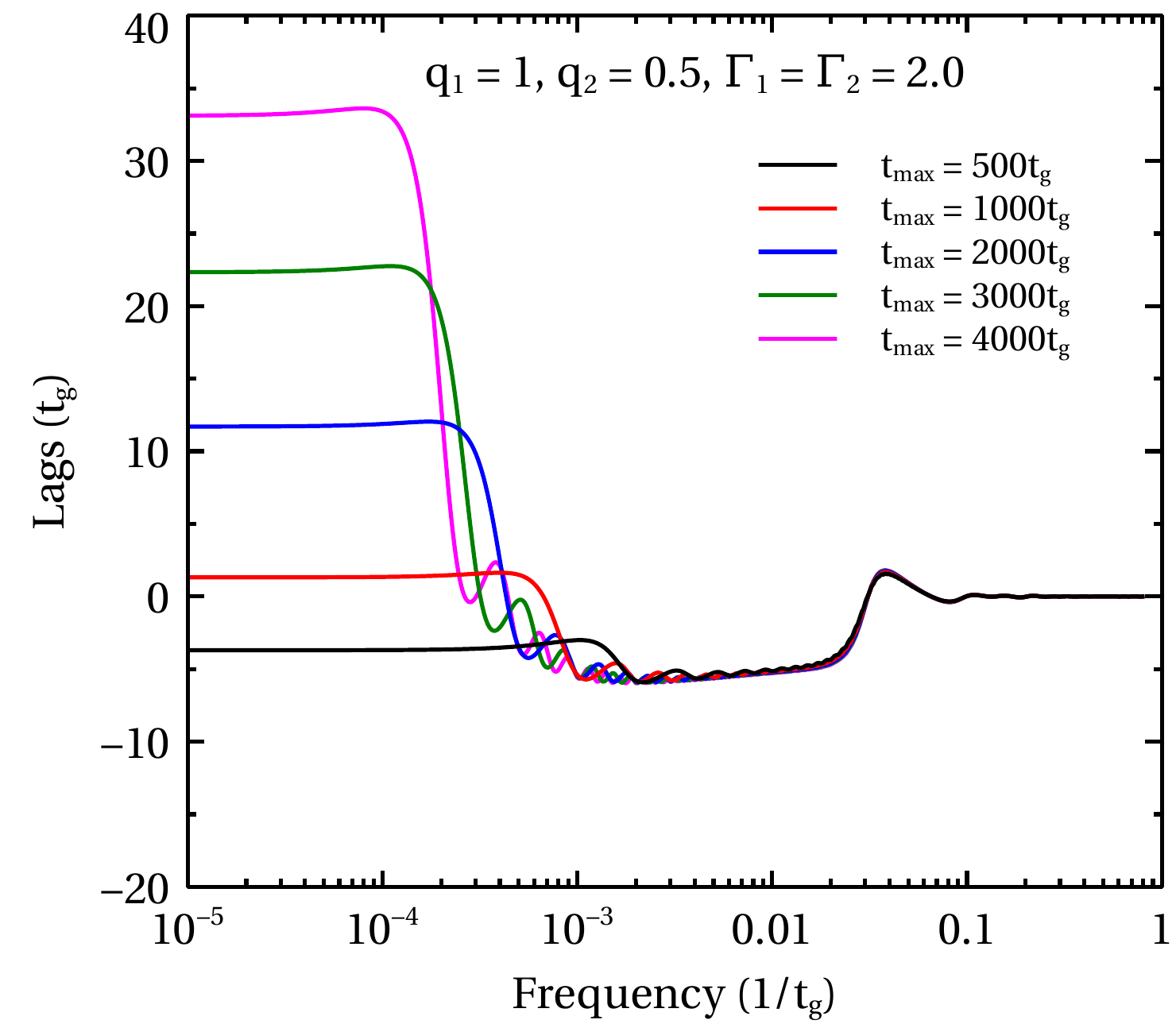}
    \caption{Frequency-dependent Fe L lags varying with $t_{\text{max}}$ when the source-response function is in the form of $ \Psi_{i}(t) \propto t^{-q_{i}} \text{exp}(-t/t_{\text{max}})$, where $q_{1}=1$, $q_{2}=0.5$, $t_{\text{shift}}=0$ and $\Gamma_{1}=\Gamma_{2}=2$.} 
    \label{p6b}
\end{figure}

We investigate how time lags change when the source- and disc-responses are combined. The reverberation parameters are fixed at $h_{1}=5r_{\text{g}}$, $h_{2}=8r_{\text{g}}$, $\theta=30^{\circ}$, $A=2$, $\xi_\text{ms} = 10^4 \text{ erg cm s}^{-1}$, $p=0$ and $B=1$. Fig.~\ref{p6b} shows the frequency-dependent Fe L lags varying with $t_{\text{max}}$ when $t_{\text{shift}}=0$ and $\Gamma_{1}=\Gamma_{2}=2$. We set $q_{1}=1$ and $q_{2}=0.5$. As mentioned before, the subscript $i=1$ and 2 refer to the parameters of the lower and the upper sources, respectively. We find that the positive, hard lags increase with increasing $t_{\text{max}}$ of the source responses and they dominate at low frequencies (i.e. varying with large amplitude on long timescales). Increasing $t_{\text{max}}$ not only increases the maximum amplitude of hard lags but also shifts their phase wrapping to lower frequencies.   

\begin{figure}
    \centerline{
        \includegraphics*[width=0.4\textwidth]{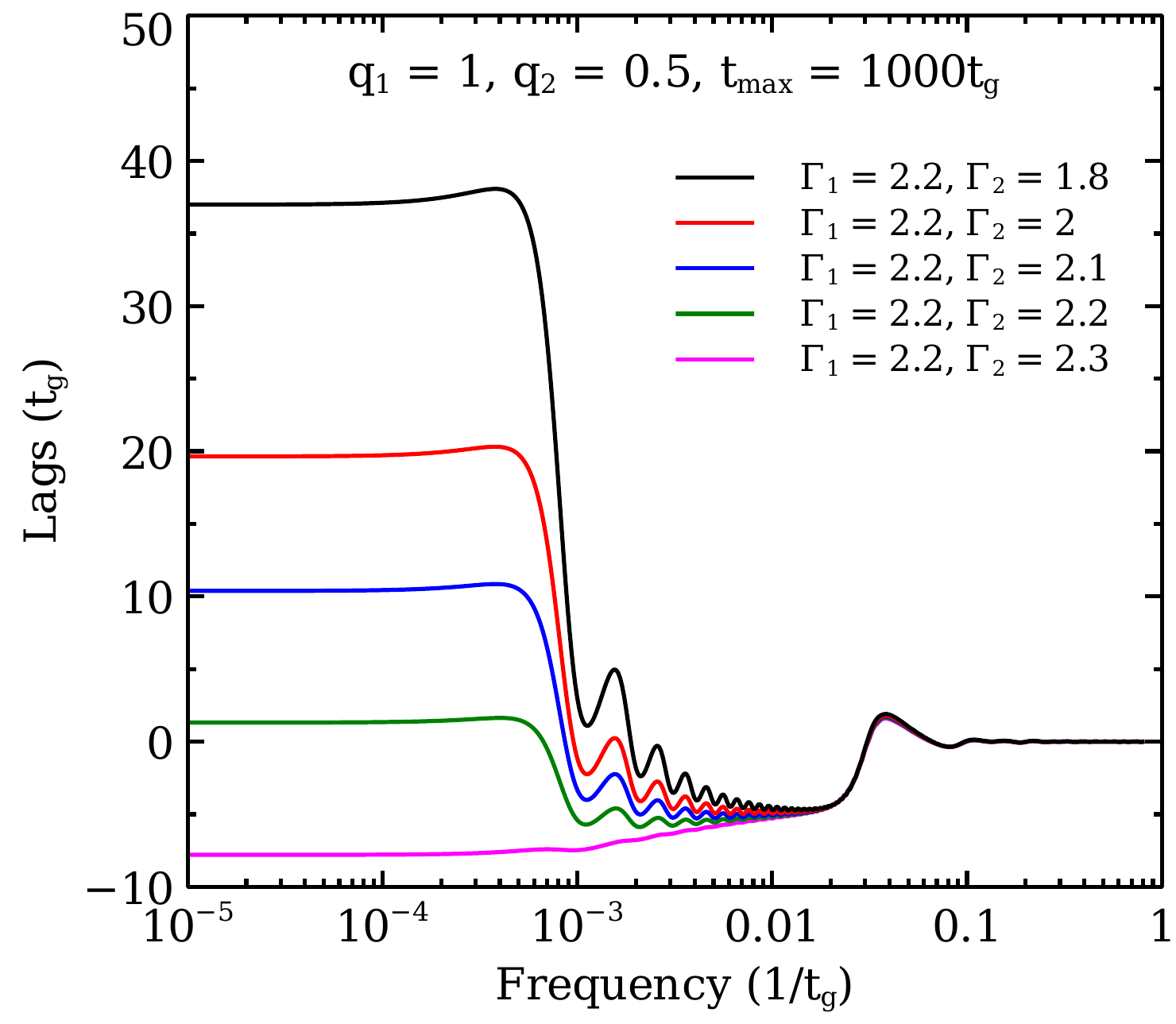}
    }
    \centerline{
        \includegraphics*[width=0.4\textwidth]{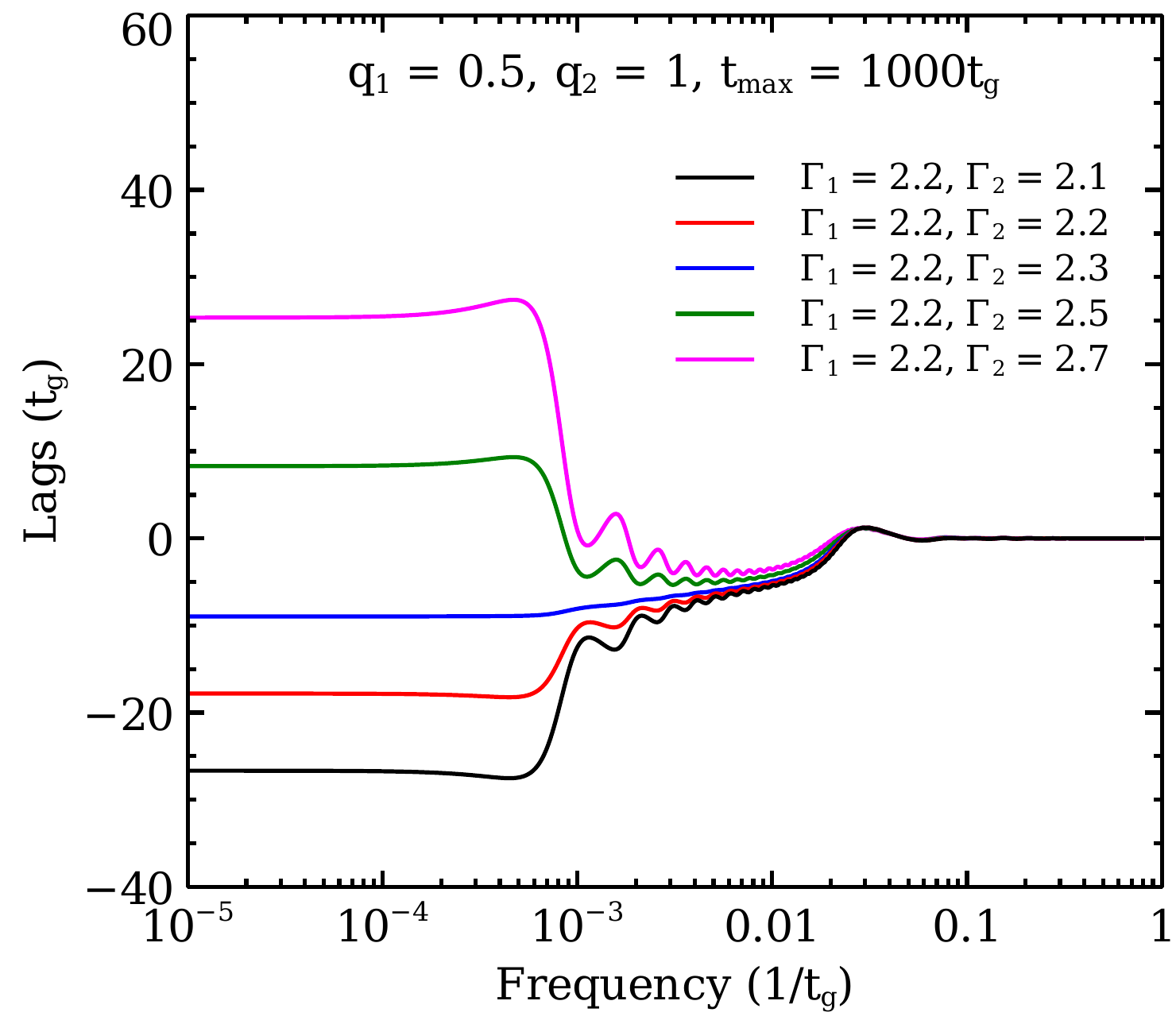}
    }
    \caption{Frequency-dependent Fe L lags varying with the photon index when $q_{1}=1$ and $q_{2}=0.5$ (top panel) and when $q_{1}=0.5$ and $q_{2}=1$ (bottom panel). The source-response function is in the form of $ \Psi_{i}(t) \propto t^{-q_{i}} \text{exp}(-t/t_{\text{max}})$ where $t_{\text{shift}}=0$ and $t_{\text{max}}=1000t_{\text{g}}$. When $\Gamma_{1} \neq \Gamma_{2}$, the model produces both high-frequency positive and low-frequency negative lags only if the harder X-ray source has lower $q$ value than that of the softer one (i.e. the harder source decays more slowly than the softer source). See text for more details.}
    \label{p6cg}
\end{figure}

How time lags vary with photon index is shown in Fig.~\ref{p6cg}. If two X-ray sources have different photon indices, the positive hard lags are produced only if the source-response function of the harder X-ray source decays more slowly than that of the softer source. In other words, the hard lags will appear on long timescales when there are more emitted photons from the harder X-ray source at late time. We are interested only in the cases when both positive hard and negative soft lags are successfully produced. In our model the choice of photon index does not change the source response function, but does contribute to the dilutions of the hard lags. The dilution effects do not change the phase wrapping frequency of reverberation lags \citep{Cackett2014,Chainakun2015}. This should also be true for the hard lags. The hard lags are scaled up if either the hard X-ray sources getting harder or the soft X-ray source getting softer. The hard lags once enhanced in this way will switch to negative lags at higher frequencies (since the phase wrapping frequency of hard lags remains the same) and thus narrow down the frequency ranges of reverberation lags. This is different to when $t_{\text{max}}$ is varied (Fig.~\ref{p6b}) because the phase wrapping frequency changes with $t_{\text{max}}$. Increasing hard lags with $t_{\text{max}}$ always broadens the frequency ranges of reverberation lags.

\begin{figure}
    \centering  
    \includegraphics*[width=70mm]{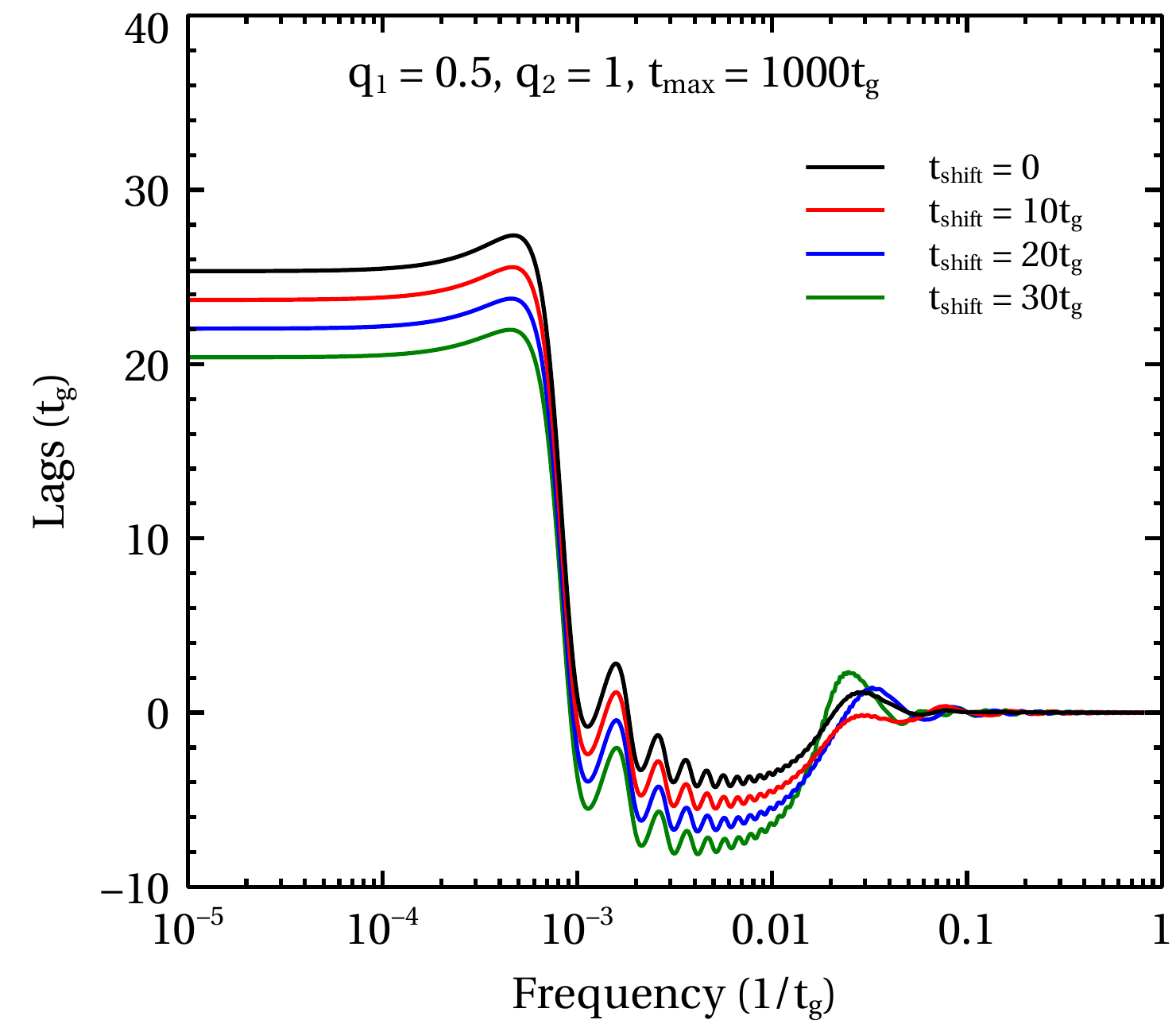}
    \caption{Frequency-dependent Fe L lags varying with $t_{\text{shift}}$ when the source-response function is in the form of $ \Psi_{i}(t) \propto t^{-q_{i}} \text{exp}(-t/t_{\text{max}})$, where $t_{\text{max}}=1000t_{\text{g}}$, $q_{1}=0.5$, $q_{2}=1$, $\Gamma_{1}=2.2$ and $\Gamma_{2}=2.7$. The changes in the lags are smaller than might be anticipated because of dilutions.} 
    \label{p6h}
\end{figure}

Fig.~\ref{p6h} shows how time lags vary with $t_{\text{shift}}$ when $\Gamma_{1}=2.2$, $\Gamma_{2}=2.7$, $q_{1}=0.5$, $q_{2}=1$ and $t_{\text{max}}=1000t_{\text{g}}$. Note that $t_{\text{shift}}$ is the time taken by fluctuations to propagate up from the lower to upper source. Since we assume the upper source is softer, increasing $t_{\text{shift}}$ means that its initial response occurs more slowly causing longer time-delays in the soft bands. The more negative soft lags increase with $t_{\text{shift}}$, the more positive hard lags decrease. However, the situation is reversed if the upper source is harder than the lower source, when larger $t_{\text{shift}}$ will lead to larger positive hard lags and smaller negative soft lags. The changes in the lags due to $t_{\text{shift}}$ are smaller than might be anticipated because of dilutions.    

\begin{figure}
    \centering
    \vspace{0.4cm}
    \includegraphics*[width=0.4\textwidth]{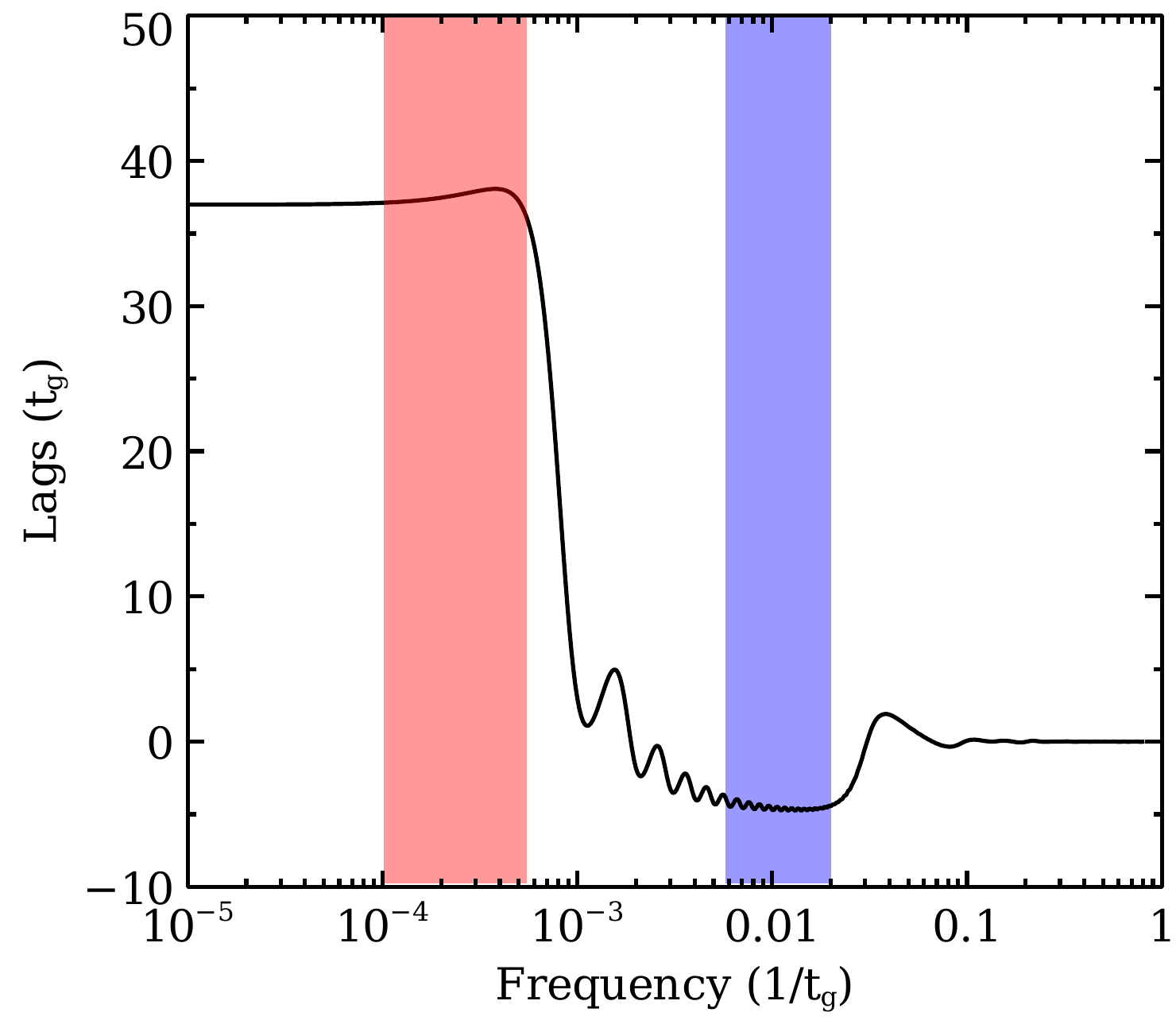}
    \centering
    \includegraphics*[width=0.4\textwidth]{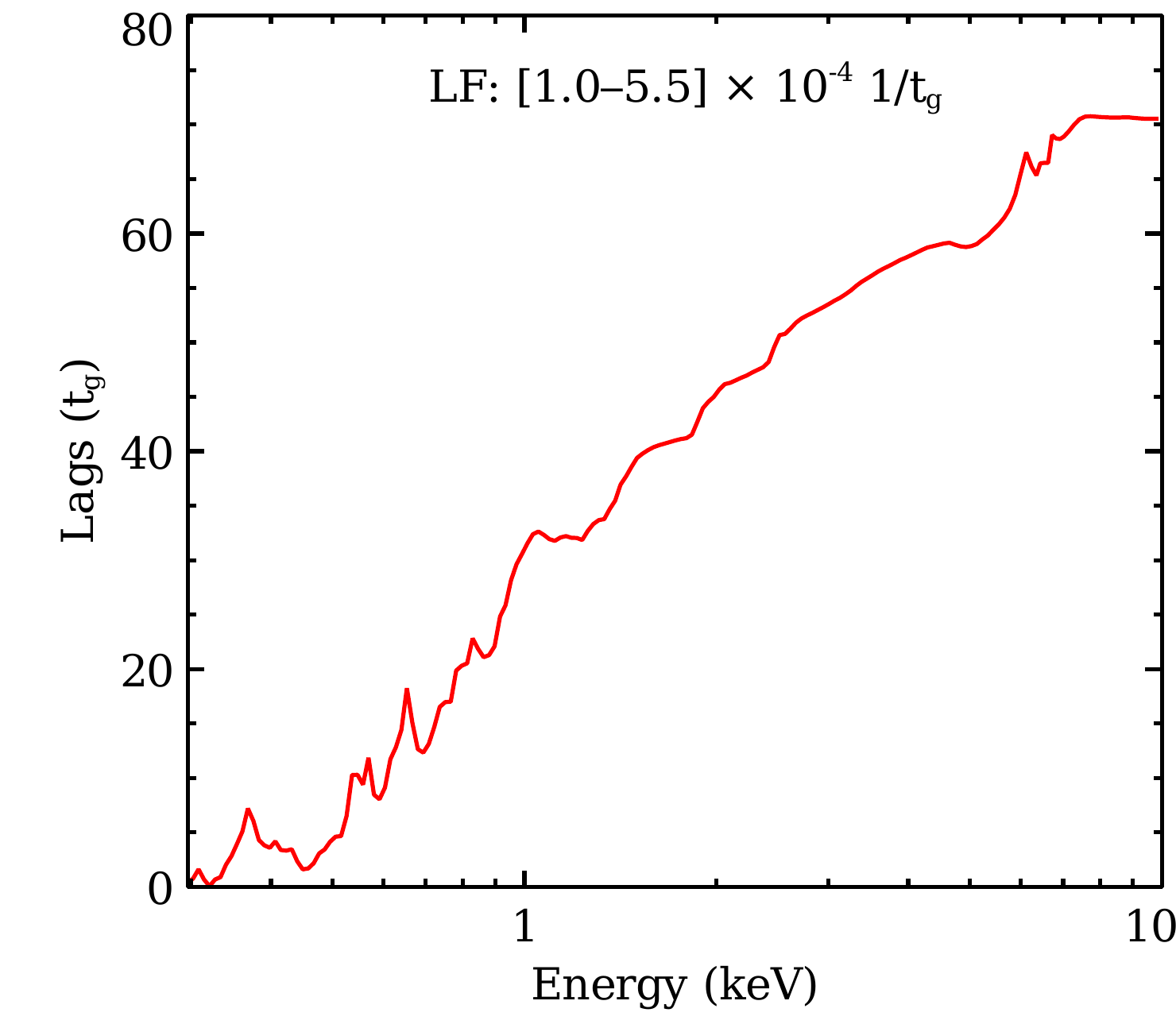}
    \centering
    \includegraphics*[width=0.4\textwidth]{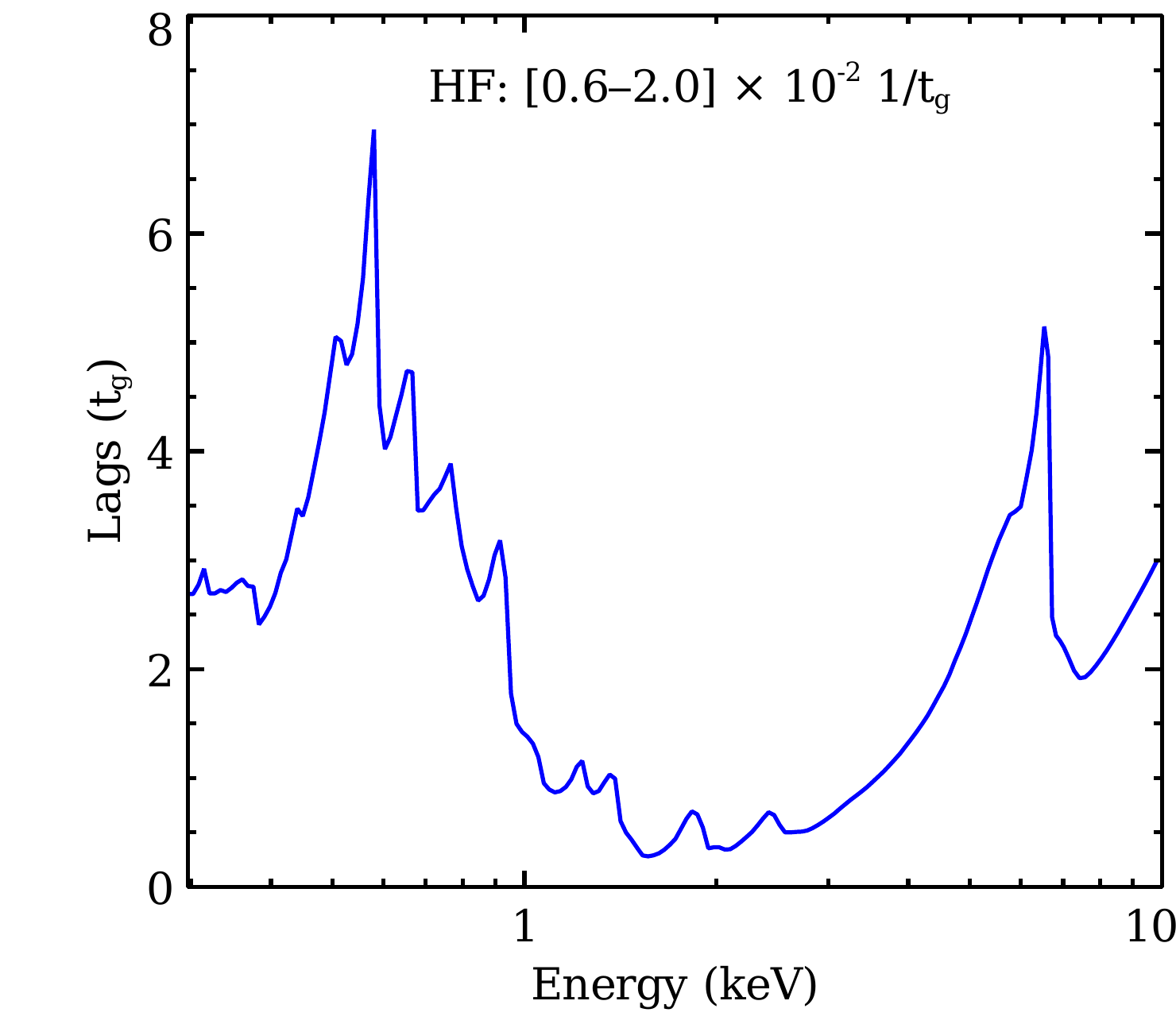}
    \caption{Top panel: Frequency-dependent Fe L lags assuming $\Psi_{i}(t) \propto t^{-q_{i}} \text{exp}(-t/t_{\text{max}})$ with $\Gamma_{1}=2.2$, $\Gamma_{2}=1.8$, $q_{1}=1$, $q_{2}=0.5$, $t_{\text{shift}}=0$ and $t_{\text{max}}=1000t_{\text{g}}$. Middle panel: corresponding lag-energy spectrum at low frequency ranges of $(1.0-5.5)\times10^{-4} 1/t_{\text{g}}$. Bottom panel: Corresponding lag-energy spectrum at high frequency ranges of $(0.6-2.0)\times10^{-2} 1/t_{\text{g}}$. }
    \label{p6def}
\end{figure}

Moreover, the energy-dependent time lags predicted by the two-blobs model are investigated. The lag-energy spectra are produced following the calculations outlined in \cite{Chainakun2016}. We focus on 0.3--10 keV band and use similar energy bins to those of {\sc reflionx} model. The energy-dependent time lags represent the relative lags of each energy bin compared to a reference band, so the bands with larger lags are lagging behind the bands with smaller lags. The reference band is selected to be the entire band excluding the band of interest, the same as when the observed lags are calculated \citep[e.g.][]{Zoghbi2011,Zoghbi2013}. Fig.~\ref{p6def} shows the frequency-dependent Fe L lags (0.3--1 vs. 2--4 keV bands) together with the corresponding lag-energy spectra extracted at low ($(1.0-5.5)\times10^{-4} 1/t_{\text{g}}$) and high ($(0.6-2.0)\times10^{-2} 1/t_{\text{g}}$) frequency ranges. The model involving the source and disc responses can generally reproduce the prominent features of time lags in AGN. At low frequencies we see the hard lags increasing with energy. Even though some of the reverberation signatures are seen at low frequencies, they are strongly dominated by the lags due to the different source responses which give the power-law trend in the lag-energy spectrum. The low-frequency lags can be featureless given the large bin-size used in producing the observed lags such as in Mrk 335 and Ark 564 \citep{Kara2013a}. At high frequencies the reverberation lags takeover and we clearly see the traditional spectral features of X-ray reverberation that relate to inner disc reflection.

\begin{figure}
    \centering  
    \includegraphics*[width=70mm]{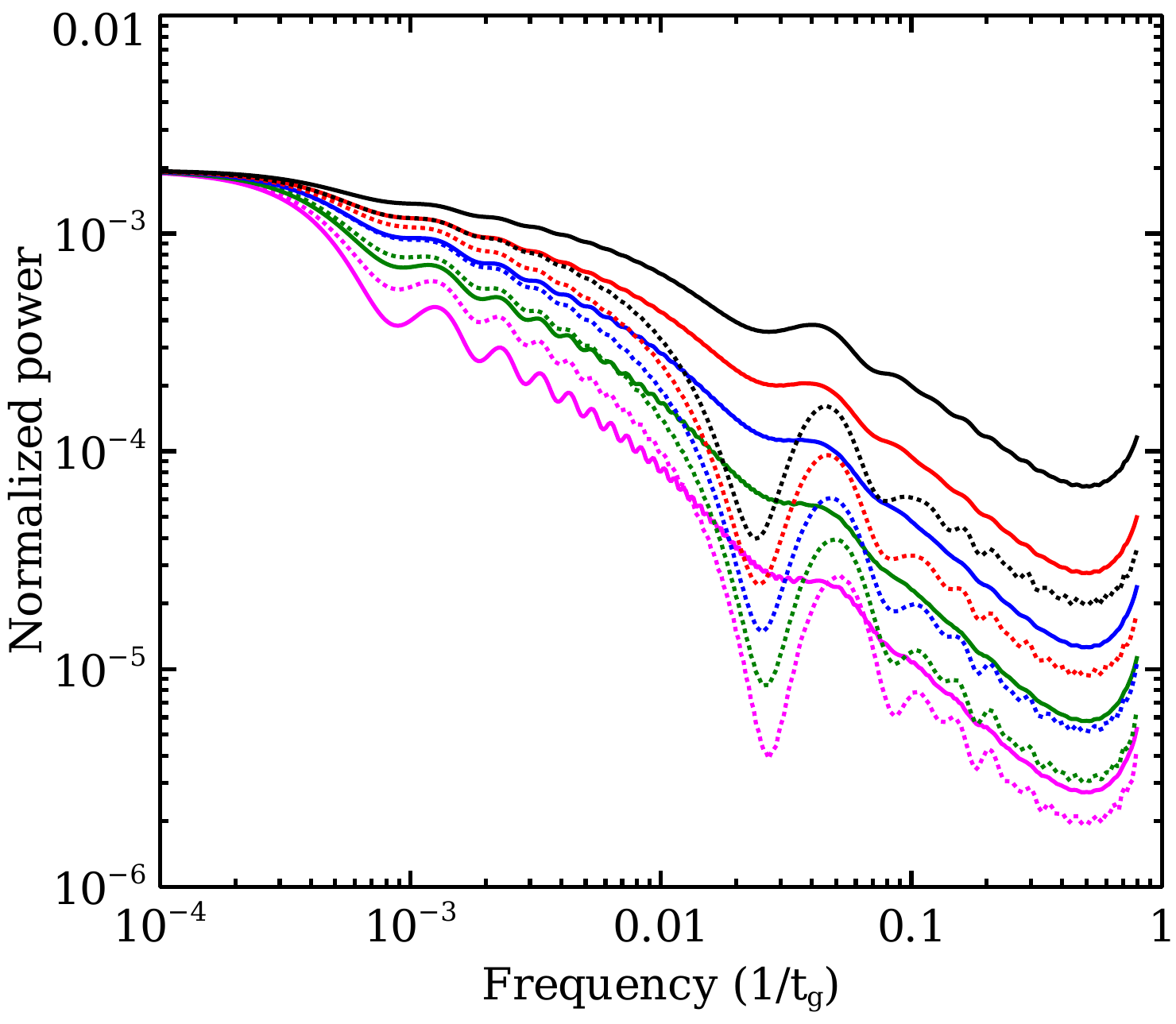}
    \caption{The PSD for the 0.3--1 keV (dotted line) and 2--4 keV (solid line) bands produced by the two-blobs model. We assume $q_{1}=1$ and $q_{2}=1.5$ (black), 1.2 (red), 1 (blue), 0.8 (green) and 0.5 (magenta). Other source parameters are $\Gamma_{1}=2.2$, $\Gamma_{2}=1.8$, $t_{\text{shift}}=0$ and $t_{\text{max}}=1000t_{\text{g}}$. Note that in this case $q_{1} < q_{2}$ and $q_{1} > q_{2}$ mean the harder source response decays faster and slower than the softer source response, respectively. The model can reproduce the traditional PSD properties seen in many AGN and X-ray binaries, in which the harder photons vary more at higher frequencies, except when the harder source response decays much slower than the softer source response (e.g. $q_{2} \lesssim 0.5$ when $q_{1}=1$ in this case). See text for more details.} 
    \label{power}
\end{figure}

Finally, it is worth checking if the predicted variability power in the soft and hard bands agrees with the observations (i.e. harder photons showing more higher frequency variability, as usually seen and explained by the propagating fluctuation model; \citealt{Kotov2001,Arevalo2006}). We use the energy-dependent light curve generated by equation~\ref{eq:obs_var} to calculate the power spectral density (PSD). The normalization is chosen to be $2\Delta t_{g} / (N \bar{a})$ \citep[e.g.][]{Miyamoto1991,Vaughan2003} where $\Delta t_{g}$ is the time bin size in graviatational units, $N$ is the number of bins and $\bar{a}$ is the mean count rate. Fig.~\ref{power} shows the PSD in the 0.3--1 keV soft and 2--4 keV hard bands produced by the two-blobs model when the first source is softer than the second source ($\Gamma_{1}=2.2$ and $\Gamma_{2}=1.8$). In this case $q_{1} < q_{2}$ and $q_{1} > q_{2}$ mean the harder source response decays faster and slower than the softer source response, respectively. We fix $t_{\text{shift}}=0$, $t_{\text{max}}=1000t_{\text{g}}$, $q_{1}=1$, and vary $q_{2}$. All PSD spectra shown in Fig.~\ref{power} exhibit a high-frequency dip whose amplitude is more prominent in the energy band that is more dominated by the reflection. The frequency where the dip is seen is energy independent. Our PSD results agree with the results of \cite{Papadakis2016} who recently investigated the effects of X-ray reprocessing on the PSD under the lamp-post assumption and found that the dip in the PSD is a common feature produced by the reverberation echo. The differences are that our total impulse response consists of not only the disc response, but also the source response that slightly affects the amplitude of the dip.

When the harder source response decays faster (Fig.~\ref{power}; black and red lines), the shorter decay timescale of the harder source results in significantly more variability power of high energy photons at high frequencies. In case of $q_{1}=q_{2}=1$ (Fig.~\ref{power}; blue lines), the observed PSD properties are obtained despite both source responses having a similar shape, providing evidence that the hard photons from the reverberation component also help generate high frequency power. We note that in this model the harder source produces both hard and soft spectra, as does the softer source. A slower decay of the harder source response then decreases the high-frequency variability power not only of the harder photons, but also of the softer photon. This is why the model can still produce more high frequency power at hard energies even when the harder source response decays slower (Fig.~\ref{power}; green lines). However, the rate of decrease of the high frequency power at soft energies is smaller than the rate at high energies. Therefore there is a limit in which the observed PSD cannot be reproduced such as when the harder source response decays much slower, e.g. $\Delta q \sim 0.5$ when $q_{1}=1$ (e.g. Fig.~\ref{power}; magenta lines). Note that the change in high-frequency variability may also depend on other model parameters and be influenced by the hard spectrum of the reverberation component, so $\Delta q \sim 0.5$ is just a rough estimate for the condition that explains the continuum hard lags in our model and also explains the observed PSD properties. Using different functions to model the source response (i.e. a response that rises instead of decaying, with a sharper rise for the harder source response and a delayed start after the softer source response) is also an option. However, since we assume each X-ray source produces both hard and soft spectra (i.e. the harder source also produces the soft spectrum), the main effects on hard lags should be due to the relative differences between the responses rather than the function itself. One possibility to always produce more high-frequency variability at harder energies is to have the harder impulse response substantially less extended in time than the softer impulse response. However, to produce hard lags our model requires the harder source response to dominate at later times, which is difficult to achieve naturally in our model if the harder source response is much shorter than the softer source response. Modelling the harder source response that is less extended in time than the softer source response requires different $t_{\rm max}$ for each X-ray source. This will further complicate the model so we do not investigate this scenario in our current paper. Instead, we show only that for small $\Delta q$ it is plausible for our initial model to mimic the X-ray properties due to the propagating fluctuations.

\section{Modelling time lags of PG~1244+026}

Previous observations showed that the lag-energy spectrum of PG 1244+026 at the frequency range of $(0.9-3.6)\times10^{-4}$ Hz shows strong Fe K lags but no soft lags at energies $< 1$~keV \citep{Kara2014}. This frequency range, however, expands over the timescales where negative (soft) and positive (hard) lags are seen \citep{Alston2014,Kara2016}. The absence of soft lags in the lag-energy spectrum may be the result of dilution by the low-frequency hard lags. This scenario could potentially be explained by the two-blobs model with different source responses as it is capable of producing the hard lags alongside the soft, reverberation lags. 

We follow the fitting technique presented in \cite{Chainakun2015} and \cite{Chainakun2016} which can be briefly described as follows. The fitting is performed in {\sc isis} \citep{Houck2000} using the {\sc subplex} method to minimise the $\chi^2$ statistic. Because of very large parameter space fitting may be very time consuming, so we choose to step through a grid of parameter values instead of interpolating between their values. We produce a global grid of model parameters to fit the lag-energy and lag-frequency spectra separately. Once a particular grid cell is found that yields the lowest $\chi^2$ value for each fit, a finer, local grid is produced around that grid cell and the fitting is repeated using these local grids to determine the best-fitting parameters.

Note that the brightness ratio is defined as $B=B_{2}/B_{1}$ where $B_{1}=1$. To minimize the model parameters we set $q_{1}=1$ because only relative difference between the two source responses should matter. The photon index of the lower source and inclination are also fixed at values in an agreement with previous studies, which are $\Gamma_{1}\sim2.2$ and $\theta \sim 30^{\circ}$ \citep{Jin2013,Kara2014}. The parameter $\Gamma_{1}$ and $\theta$ have small effects on reverberation lags under the lamp-post assumption \citep{Cackett2014,Chainakun2016}, especially when the bin sizes are large. However, the relative difference of the photon index between two X-ray sources significantly change the overall time lags in the two-blobs model, so while $\Gamma_{1}$ is fixed at 2.2, $\Gamma_{2}$ is allowed to vary. When fitting the data using the global grid, we fix $B=1$ and find the upper source is at $h_{2}\sim11r_{\text{g}}$. However, the best-fitting models suggest the very small amounts of reflection from the upper source. When expanding the local grid, we then neglect the reflection component of that source, allow $B$ to vary and choose to determine $h_{2}$ from the constrained parameter $B$, which will be discussed in the next Section. This helps speed up the computations in producing the local-grid cells. 

\begin{figure}
    \centering  
    \includegraphics*[width=90mm]{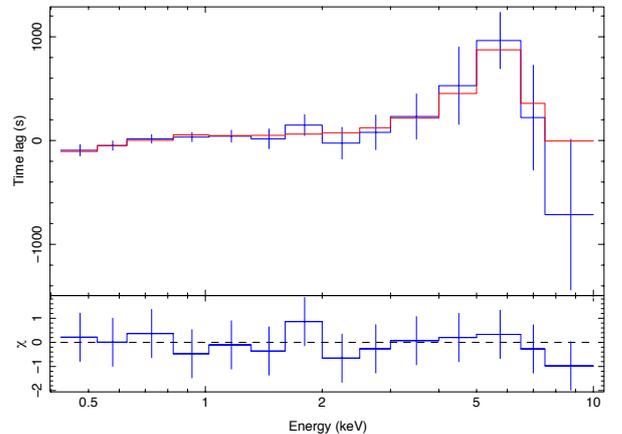}
    \vspace{-0.7cm}
    \caption{Data and residuals from fitting the two-blob model to the lag-energy spectrum of PG~1244+026, in the frequency range of $(0.9-3.6)\times10^{-4}$ Hz. } 
    \label{fit-lag-e}
\end{figure}

\begin{figure}
    \centering  
    \includegraphics*[width=90mm]{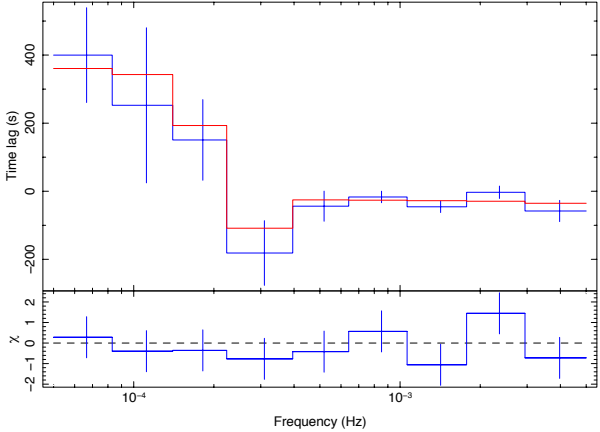}
    \vspace{-0.7cm}
    \caption{Data and residuals from fitting the two-blob model to the lag-frequency spectrum between 0.3--1 and 1--4 keV bands of PG~1244+026. } 
    \label{fit-lag-f}
\end{figure}

The fitting results for lag-energy and lag-frequency spectra are presented in Fig.~\ref{fit-lag-e} and Fig.~\ref{fit-lag-f}, respectively. The constrained parameters are listed in Table~\ref{best-para}. The errors are quoted at $1\sigma$ confidence intervals around the best-fitting values ($\Delta\chi^{2}=1.0$ for each parameter of interest). There is a degeneracy between $\xi_\text{ms}$ and $p$ so when calculating the errors we fix $p=2.4$ which is its best fitting value and allow other parameters to vary independently. We find the fitting results from lag-energy and lag-frequency models are quite comparable, despite them being independently fit to the data.   

\begin{table}
    \begin{tabular}{lll}
        \hline
        \multirow{2}{*}{Parameter} & \multirow{2}{*}{Lag-energy fits} &  \multirow{2}{*}{Lag-frequency fits}\\  
        \\ \hline \vspace{0.1cm}
        $h_{1} (r_{\text{g}})$ & $6.0^{+2.0}_{-1.0}$ & $8.0^{+2.0}_{-2.0}$ \\ \vspace{0.1cm}
        $\theta$ ($^{\circ}$) & $30^{f}$ & $30^{f}$ \\ \vspace{0.1cm}
        $\Gamma_{1}$  & $2.2^{f}$ & $2.2^{f}$ \\ \vspace{0.1cm}
        $\Gamma_{2}$  & $3.5^{+0.3}_{-0.4}$ & $3.0^{+0.5}_{-0.2}$ \\ \vspace{0.1cm}
        $A$  & $5^{+5}_{-3}$ & $5^{+5}_{-3}$\\ \vspace{0.1cm}
        log $\xi_\text{ms} (\text{ erg cm s}^{-1})$ & $3.4^{+0.4}_{-0.2}$ & $3.4^{+0.8}_{-0.4}$ \\  \vspace{0.1cm}
        $B$ & $2.7^{+0.5}_{-1.0}$ & $2.0^{+1.0}_{-0.4}$\\ \vspace{0.1cm}
        $q_{2}$ & $1.1^{+0.1}_{-0.1}$ & $1.1^{+0.1}_{-0.1}$\\ \vspace{0.1cm}
        $t_\text{shift}$ $(t_{\text{g}})$ & $0^{+3}_{--}$ & $1^{+1}_{-1}$\\ \vspace{0.1cm}
        $t_\text{max}$ $(t_{\text{g}})$ & $500^{+1500}_{-400}$ & $500^{+1500}_{-400}$\\ \vspace{0.1cm}
        log $M (M_\odot)$ &  $7.34^{+0.10}_{-0.08}$ & $7.27^{+0.05}_{-0.04}$ \\  
        \hline
        $\chi^{2} / \text{d.o.f.}$ & 0.79 & 1.14 \\
        \hline
    \end{tabular}
    \caption{The best-fitting parameters constrained by the model. The parameters and their values for the fits of lag-energy and lag-frequency spectra are listed in Columns 1, 2 and 3 respectively. The errors correspond to $1\sigma$ confidence level for one parameter of interest and are estimated by linear interpolation if necessary. The superscript $f$ indicates the parameters which are fixed. Note that we set $q_{1}=1.0$ and $B=B_{2}/B_{1}$ where $B_{1}=1.0$. The two fits were performed independently. ``--'' indicates that the error range cannot be estimated because of the finite extension of the grid cells. See text for more details.
    \label{best-para}}
\end{table}

\section{Discussion}

Even though we carry out this work through the perspective of timing analysis alone, it should be noted that there are degeneracies in the spectral fitting of PG~1244+026 \citep{Jin2013}. Some of our parameters ($\Gamma_{1}$ and $\theta$) are fixed at the values approximately in an agreement with the relativistic reflection models of \cite{Jin2013} and \cite{Kara2014}. Those models assumed a constant ionization disc whose ionization parameter is $\xi \sim 500$ $\text{ erg cm s}^{-1}$ and $\xi \sim 1,000$ $\text{ erg cm s}^{-1}$, respectively. Here we consider a disc with an ionization gradient which is found to be moderately ionized at the innermost region, $\xi_\text{ms} =2,500\text{ erg cm s}^{-1}$, which is not too high so the disc annuli not far from the centre where the relativistic effects are still important should possess those reported inonization values. Their iron abundance was constrained at $A\sim2$, which is the low end of $1\sigma$ confidence level for our best-fitting iron abundance. 

The central mass of PG~1244+026 has been measured using various techniques. Using optical reverberation, the black hole mass related to the Broad Line Region (BLR) size was found to be $M\sim 4.8 \times 10^6 M_\odot$ \citep{Vestergaard2006}. They assumed there is no radiation force compensating for gravitational attraction that pulls the BLR clouds. Taking into account the effects of radiation pressure from ionizing photons, a significantly larger mass of $M\sim 1.8 \times 10^7 M_\odot$ was found to effectively pull the BLR clouds \citep{Marconi2008}. A study of the correlation between the excess variance (i.e. estimator of the intrinsic source variance) and black hole mass suggested a mass of $M\sim 0.5-1.5 \times 10^7 M_\odot$ \citep{Ponti2012}. The averaged excess variance was also derived independently by \cite{Done2013} who found the correlated mass range of $M\sim 0.2-2.0 \times 10^7 M_\odot$. Moreover, the broad-band Spectral Energy Distribution (SED) analysis carried out by \cite{Jin2013} put the mass at $M\sim 1.6 \times 10^7 M_\odot$. Fitting the X-ray reverberation Fe K lags revealed the central mass to be $M\sim 1.3 \times 10^7 M_\odot$ \citep{Kara2014}. Using two-blobs model, we find the black hole mass of $\sim  1.8-2.2\times 10^7 M_\odot$, around the high end of values found by previous methods.     

Our best-fitting models place the $1^{\text{st}}$ source at $\sim 6-8r_{\text{g}}$, but the source height found by \cite{Kara2014} using the x-ray reverberation under the lamp-post scheme was $\sim5r_{\text{g}}$. Although different geometries are assumed, the predicted lags should reflect the averaged light-crossing time between the source(s) and the disc, regardless of what the source geometries are. The distance given by $h_{1}$ sets the lower limit of intrinsic reverberation lags in the two-blobs model (see Fig.~\ref{p1}). Therefore our source geometry should always produce larger intrinsic reverberation lags compared to the source geometry of \cite{Kara2014}. This implies that the dilution in our model is larger so that both models agree with the same observed reverberation lags. 

How the dilution effects are taken into account is very important \citep[e.g.][]{Uttley2014}. The timing model of \cite{Kara2014} relied on the calculation of \cite{Cackett2014} where the lag-energy spectrum was estimated by tracing the Fe K photons and was produced by X-ray reverberation alone. Here the photons of all X-ray energies are traced so the Fe K lags are formed not only by the Fe K photons but also by photons of other energies being red-shifted or blue-shifted. Both direct and reflection photons are included in each energy band so that the full effects of dilution are taken into account. In addition to dilution effects due to the relative contribution of the direct and reflection components, our reverberation lags are further diluted by the positive hard lags due to the different source responses. As a result, our model is subject to a large amount of dilution which places the sources at higher position on the symmetry axis of the black hole.          

Since the model requires only a small amount of X-ray reflection from the $2^{\text{nd}}$ source, we interpret this source as moving upwards rapidly so that its emission is beamed away from the accretion disc. In order to estimate the location of the $2^{\text{nd}}$ X-ray source, we evaluate the brightness ratio, $B$, using the standard light bending model \citep[e.g.][]{Miniutti2004}. The observed continuum flux from the axial source with fixed intrinsic luminosity is expected to increase with the source height. According to this, if two X-ray sources have the same intrinsic luminosity, our parameter $h_{1}\sim 6-8r_{\text{g}}$ and $B\sim 2-2.7$ implies $h_{2} \sim 11r_{\text{g}}$. Nevertheless, this is a rough estimate and if its intrinsic luminosity is relatively higher or lower the location of the $2^{\text{nd}}$ source can be lower or higher than $11r_{\text{g}}$, respectively.

PG~1244+026 is one of a few Seyfert galaxies so far that shows strong reverberation signatures but the high frequency lag-energy spectrum has no soft-excess and no dip at $\sim 3$ keV. \cite{Chainakun2016} suggested that the dip at $\sim 3$ keV (i.e. the 3 keV band leads the adjacent bands) can be reproduced under the lamp-post assumption with a disc with an ionization gradient while \cite{Wilkins2016} showed that the X-ray propagation up the black hole rotation axis through a vertically collimated corona help enhance this dip. The absence of a clear 3 keV dip is possibly because either the central vertically extended corona does not form or it does form but there are no propagating fluctuations. It is, however, ambiguous for PG~1244+026 as we have not seen its soft excess lags so the soft bands may not be diluted in the same. If the lags of PG~1244+026 are more diluted in the softer bands than in the 3 keV band, the non-detection of the 3 keV dip is possible even though the geometry and production mechanism support this dip.

Nevertheless, our lag-frequency and lag-energy fitting results are self-consistent and both agree on the very small value of $t_\text{shift}$. Given the acceptable range of source separation distance of $\sim 2-7r_{\text{g}}$ and the high end of the acceptable range for $t_\text{shift}\sim 3r_{\text{g}}$, our model is still open to possibility of propagation between the sources at near the speed of light. However, as pointed out by \cite{Wilkins2016}, the propagation speed of $<0.01c$ is required in order to produce the clear 3~keV dip. Thus our upper limit on $t_\text{shift}$ is sufficient to rule out the possibility that PG~1244+026 possesses the geometry that produces a 3~keV dip which is obscured due to the dilution effects. The intrinsic time lags themselves do not have a clear 3~keV dip. Also, the downward-propagation scenario was tested but the fits were worse. We conclude that only the near speed-of-light propagation up through the vertically extended X-ray sources of PG~1244+026 is allowed during this observation.

A global look through all variable and well-observed Seyfert galaxies available in the \emph{XMM-Newton} archive to date found high frequency reverberation lags in $\sim 50$ per cent of sources and low frequency hard lags in $\sim 85$ per cent of sources \citep{Kara2016}. We show that the two-blobs model with different source responses can replicate both kinds of time lags. We cannot make a strong claim on $t_\text{max}$ because the parameters $q_{1}$ and $q_{2}$ are slightly different and hence the shapes of corresponding response functions are quite close. This is why the error bars for $t_\text{max}$ are large. Precise constraint of $t_\text{max}$ requires higher quality data that have smaller errors on the low frequency lags. The source parameters in our model make a huge contribution to the low-frequency hard lags. There is also evidence that the presence of a warm absorber may affect the lags on long timescales \citep{Silva2016} giving more uncertainty in constraining the source parameters. 

Although the impulse source responses are modelled using a power-law, other choices are possible to produce the hard lags as long as the softer source response dominates at the beginning and the harder source response dominates towards the end. More high-frequency variability at harder energies will always be produced if we assume the harder source response is substantially less extended in time than the softer source response. This, however, requires additional parameters for different cut-off response time that will make the model more complicated and hence is beyond the scope of this paper. Despite this, we have already shown in Section 2.2 that a simple power-law as a choice of the source response is plausible enough and can produce more high frequency power at hard energies even when the harder source response decays slower, but within a limited range, e.g., $\Delta q \lesssim 0.5$. This is because the hard photons from the reverberation components show significant high-frequency variability that can counteract the power loss by the harder photons varying less on shorter timescales due to relatively slower the decay of the harder source response. It should be noted that a slower decay of the harder source response in our model leads to soft photons showing less high-frequency variability as well.


Our model does not capture the lags that may arise from the disc blackbody emission. However, the lags associated with the soft excess are complex and may depend upon the choice of reflection model \citep{Chainakun2015}. The interpretation of soft excess is uncertain and the blackbody emission, if exists, is therefore plausible to be treated as a non-variable component. \cite{Jin2013} reported the uncorrelated variability of the soft excess in this source. The spectral models with a separate soft excess component having its own lag were also proposed by \cite{Alston2014}. This separate lag from another soft component can alternatively be explained by the second X-ray source in our model that produces only the soft continuum, affecting the lags more in the soft bands.

\begin{figure}
    \centering
    \includegraphics*[width=70mm]{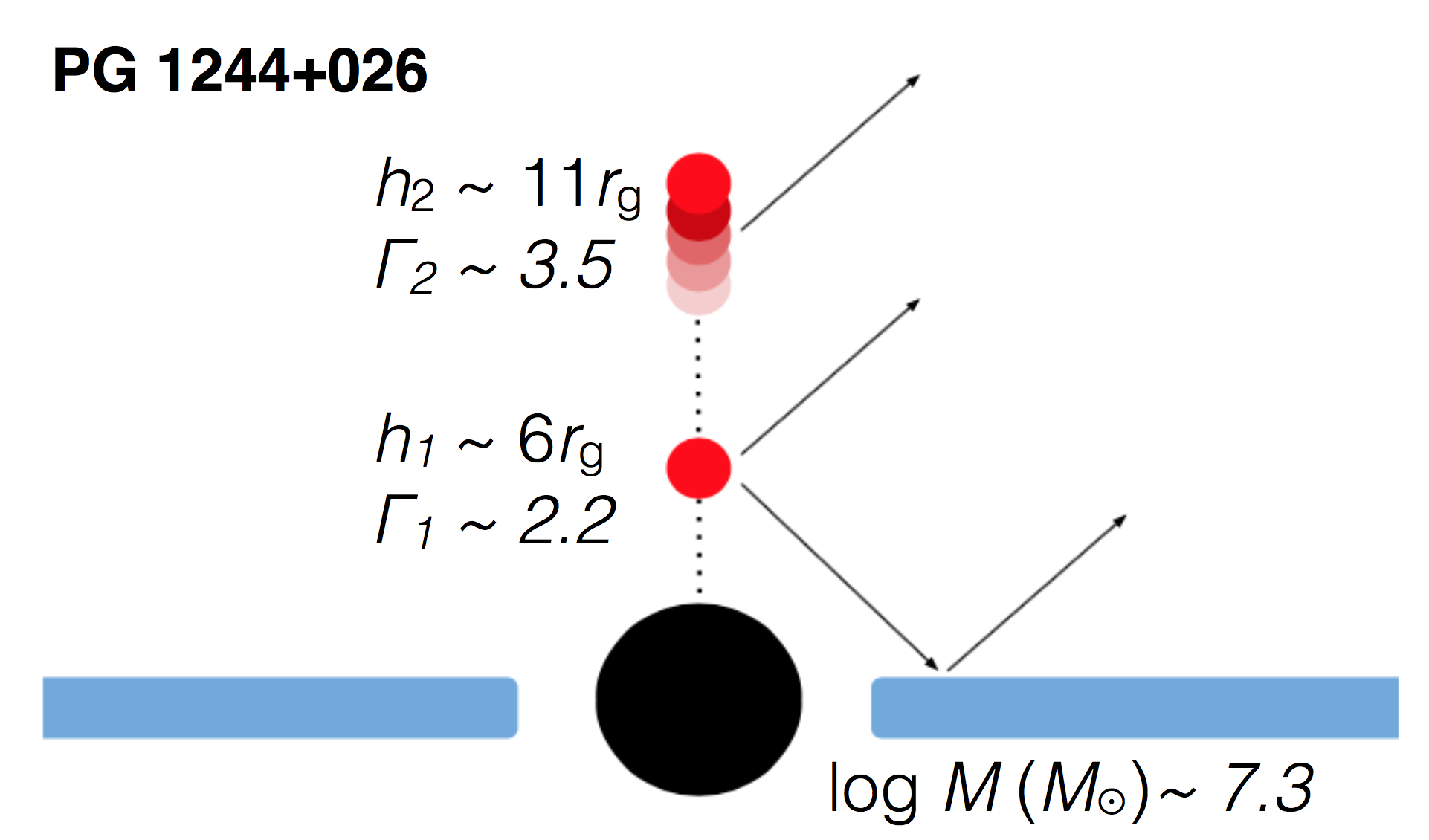}
    \caption{Sketch of the geometry of PG~1244+026 constrained by two-blobs model. }
    \label{geometry}
\end{figure}

\begin{figure}
    \centering  
    \vspace{-1.0cm}
    \includegraphics*[width=70mm]{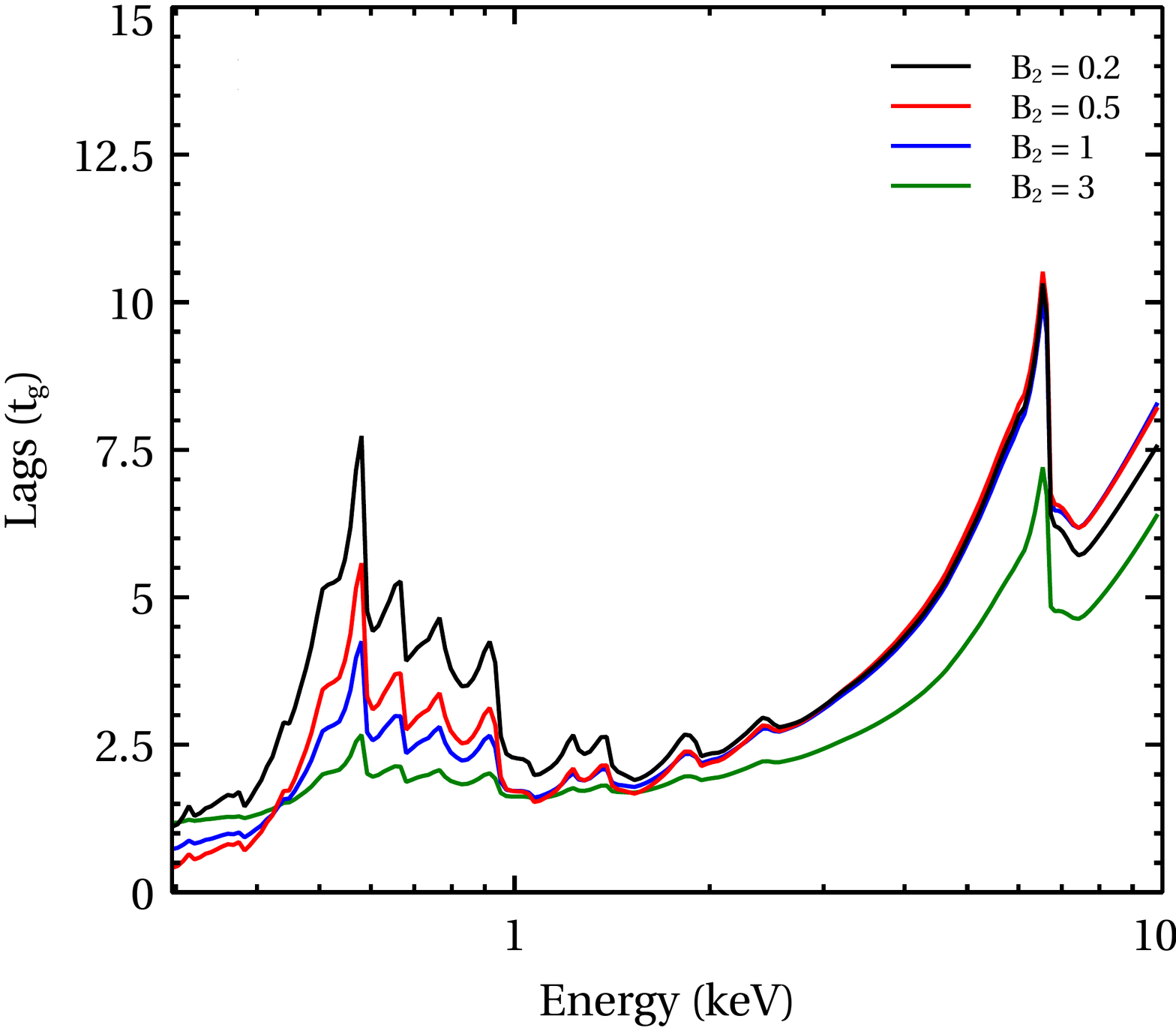}
    \vspace{-2.0cm}
    \caption{Effects of changing $B_{2}$ on the high-frequency lag-energy spectra when the $2^{\text{nd}}$ source is moving away very fast. To highlight and clarify the features we set $\Gamma_{1}=2.2$, $\Gamma_{2}=3$. As the $2^\text{nd}$ softer source gets brighter there is more dilution in the soft band.}
    \label{toy_model}
\end{figure}

Fig.~\ref{geometry} shows the sketch of the geometry of PG~1244+026 constrained by two-blobs model. If the $2^{\text{nd}}$ source moves away very fast and does not produce back-scattered X-rays from the disc, the absence of its reflection flux makes its continuum flux to be a contamination component diluting the lags of the $1^{\text{st}}$ source. To prove this, we produce a toy model where the $2^{\text{nd}}$ source is softer, producing only the direct continuum and the parameter $B$ is varied. Predicted time lags of this toy model are shown in Fig.~\ref{toy_model}. The dilution effects seem to be stronger in the soft band than in the Fe K band, which is what we expected since the moving source is producing more soft X-ray continuum. If the $2^\text{nd}$ source is brighter, the lags are further diluted especially in the band its continuum significantly contributes to. On the other hand, the presence of the $2^\text{nd}$ source is irrelevant when it is very faint ($B \ll 1$) and the corresponding time lags should be similar to the lamp-post case. The comparison between time lags under the standard and modified geometries (including the source geometry of PG~1244+026 and the geometry when the observer and the disc see different part of the corona) is also presented in the Appendix. Nevertheless, it is important to note that the two-blobs model can produce either smaller or larger reverberation lags comparing to the lamp-post case where the X-ray source is placed at $h_{1}$. Including the second, upper source will increase intrinsic reverberation lags only if the upper source produces both direct and reflection spectra (Fig.~\ref{p1}). If the upper source emits only the direct continuum, it will dilute the intrinsic lags produced by the source at $h_{1}$.

\begin{figure}
    \centering
    \includegraphics*[width=90mm]{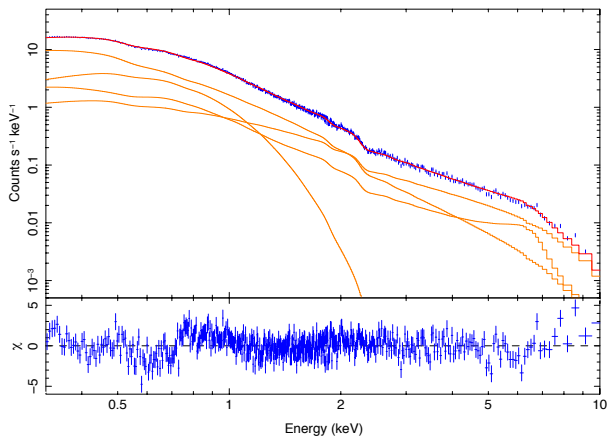}
    \vspace{-0.7cm}
    \caption{Data and residuals when the time-averaged spectrum of PG~1244+026 is fitted with the model mean-spectrum reproduced using the best-fitting parameters constrained by timing models. Most of the model parameters were frozen at values consistent with the timing data fits. The data and the model are presented in blue and red, respectively. The individual components shown in orange consist of two power-law spectra (from two sources), one blurred reflection spectrum (from only the lower source) and a blackbody component that contributes significant flux at energies $<1$~keV. } 
    \label{mean_spectrum}
\end{figure}

The two-blobs models provides good fits to the data even though we do not interpolate the parameter values between grid cells. Fig.~\ref{mean_spectrum} shows the time-averaged spectrum produced using the parameters from our best-fitting timing models. Additional blackbody emission and warm absorber are required. Both are assumed to have no contribution to the time lags. The spectral fits are acceptable ($\chi^{2} / \text{d.o.f.}=1.63$) but they can be further improved by interpolating the models, which unfortunately is very time consuming with our approach. Our lag-energy models seem to over-fit the data ($\chi^{2} / \text{d.o.f.} < 1.0$). This is unavoidable if we begin to use a physically motivated model which definitely requires many parameters to explain time lags that have relatively small number of data points. According to this, better quality of data is essential in order to fit the timing profiles under the complex, realistic geometry beyond the lamp-post one. Longer observations from \emph{XMM-Newton} or future observations made by \emph{Athena} \citep[e.g.][]{Dovciak2013} will deliver a better understanding of geometries and properties of the complex X-ray sources in AGN.

\section{Conclusion}

We present a theoretical model to produce the X-ray time lags in AGN using two X-ray point sources. We show that the two-blobs model can reduce to the lamp-post case (where there is only one X-ray source at $h_{1}$) when the upper source at $h_{2}$ is relatively very faint. The reverberation lags can either increase or decrease subjected to the existence of the upper X-ray source. The distance given by $h_{1}$ always sets the lower limit of intrinsic reverberation lags in the two-blobs model. If the upper source produces both direct and reflection components, the reverberation lags will increase due to the longer time delays of photons with the longer travel distance $h_{2}$. Contrarily, if the upper source produces only the power-law continuum, the reverberation lags will be diluted. Furthermore, we assume the variations of two sources that may be triggered by the same mechanism (e.g. perturbations due to accretion rate fluctuations propagating through the disc). The positive hard and negative soft lags in the lag-frequency spectrum can be produced by assuming different source responses for two X-ray sources. The model can also reproduce the power-law profile for the low-frequency lag-energy spectrum (e.g., harder band lagging the softer bands) and imprint the reverberation signatures in the high-frequency lag-energy spectrum. These characteristic features are observable in many AGN \citep[see][for the global features of X-ray time lags in AGN]{Kara2016}. Therefore, in principle, the properties of time lags in most AGN should be explained by using only two X-ray sources, even though the full modelling of extended source \citep{Wilkins2016} is very important. However, the models beyond the lamp-post assumption have many parameters so fitting them to the data is still very computationally intensive. 

In this paper we fit the timing data of PG~1244+026, the AGN that showed complex variability in the soft bands \citep[e.g.][]{Jin2013,Kara2014}. Our best-fitting results place two X-ray sources at $h_{1}\sim6r_{\text{g}}$ and $h_{2}\sim11r_{\text{g}}$. These sources respond to the primary variation in slightly different way ($q_{1}$ and $q_{2}$ are different). The fitting of lag-frequency, lag-energy and mean-spectrum all approximately agree on the same source geometry even though they are not simultaneously fit. The upper source produces small amounts of reflection X-rays suggesting a feasible geometry of relativistic jet where the emission photons are beamed away from the disc. Moreover, the fluctuations propagating upwards between the two sources are allowed for this AGN but only at near the speed of light (far beyond $0.01c$). This rules out the possibility that the absence of the 3~keV dip is due to very strong dilution in the soft bands. The intrinsic properties of the disc and the corona themselves do not support this dip. The two-blobs model provides an independent approach to map out the sources and their relative variability. We include the effects of ionization gradients in the disc and consider the full reflection and continuum components so that all energy bands are self-consistently diluted. This model is a logical step towards realistic modelling of time lags from a full-extended source which is planned for the future.


\section*{Acknowledgements}
This work was carried out using the computational facilities of the Advanced Computing Research Centre, University of Bristol. PC thanks the University of Bristol for a Postgraduate Research Scholarship. We thank Erin Kara for providing the observational timing data of PG~1244+026. We thank the anonymous referee for carefully reading the paper and the very helpful suggestion that we consider the power spectral density at different energies.

\bibliographystyle{mnras}
\bibliography{bibdata}

\begin{thebibliography}{}
\makeatletter
\relax
\def\mn@urlcharsother{\let\do\@makeother \do\$\do\&\do\#\do\^\do\_\do\%\do\~}
\def\mn@doi{\begingroup\mn@urlcharsother \@ifnextchar [ {\mn@doi@}
  {\mn@doi@[]}}
\def\mn@doi@[#1]#2{\def\@tempa{#1}\ifx\@tempa\@empty \href
  {http://dx.doi.org/#2} {doi:#2}\else \href {http://dx.doi.org/#2} {#1}\fi
  \endgroup}
\def\mn@eprint#1#2{\mn@eprint@#1:#2::\@nil}
\def\mn@eprint@arXiv#1{\href {http://arxiv.org/abs/#1} {{\tt arXiv:#1}}}
\def\mn@eprint@dblp#1{\href {http://dblp.uni-trier.de/rec/bibtex/#1.xml}
  {dblp:#1}}
\def\mn@eprint@#1:#2:#3:#4\@nil{\def\@tempa {#1}\def\@tempb {#2}\def\@tempc
  {#3}\ifx \@tempc \@empty \let \@tempc \@tempb \let \@tempb \@tempa \fi \ifx
  \@tempb \@empty \def\@tempb {arXiv}\fi \@ifundefined
  {mn@eprint@\@tempb}{\@tempb:\@tempc}{\expandafter \expandafter \csname
  mn@eprint@\@tempb\endcsname \expandafter{\@tempc}}}

\bibitem[\protect\citeauthoryear{{Alston}, {Done}  \& {Vaughan}}{{Alston}
  et~al.}{2014}]{Alston2014}
{Alston} W.~N.,  {Done} C.,   {Vaughan} S.,  2014, \mn@doi [\mnras]
  {10.1093/mnras/stu005}, \href
  {http://adsabs.harvard.edu/abs/2014MNRAS.439.1548A} {439, 1548}

\bibitem[\protect\citeauthoryear{{Ar{\'e}valo} \& {Uttley}}{{Ar{\'e}valo} \&
  {Uttley}}{2006}]{Arevalo2006}
{Ar{\'e}valo} P.,  {Uttley} P.,  2006, \mn@doi [\mnras]
  {10.1111/j.1365-2966.2006.09989.x}, \href
  {http://adsabs.harvard.edu/abs/2006MNRAS.367..801A} {367, 801}

\bibitem[\protect\citeauthoryear{{Balbus} \& {Hawley}}{{Balbus} \&
  {Hawley}}{1991}]{Balbus1991}
{Balbus} S.~A.,  {Hawley} J.~F.,  1991, \mn@doi [\apj] {10.1086/170270}, \href
  {http://adsabs.harvard.edu/abs/1991ApJ...376..214B} {376, 214}

\bibitem[\protect\citeauthoryear{{Cackett}, {Zoghbi}, {Reynolds}, {Fabian},
  {Kara}, {Uttley}  \& {Wilkins}}{{Cackett} et~al.}{2014}]{Cackett2014}
{Cackett} E.~M.,  {Zoghbi} A.,  {Reynolds} C.,  {Fabian} A.~C.,  {Kara} E.,
  {Uttley} P.,   {Wilkins} D.~R.,  2014, \mn@doi [\mnras]
  {10.1093/mnras/stt2424}, \href
  {http://adsabs.harvard.edu/abs/2014MNRAS.438.2980C} {438, 2980}

\bibitem[\protect\citeauthoryear{{Chainakun} \& {Young}}{{Chainakun} \&
  {Young}}{2012}]{Chainakun2012}
{Chainakun} P.,  {Young} A.~J.,  2012, \mn@doi [\mnras]
  {10.1111/j.1365-2966.2011.20105.x}, \href
  {http://adsabs.harvard.edu/abs/2012MNRAS.420.1145C} {420, 1145}

\bibitem[\protect\citeauthoryear{{Chainakun} \& {Young}}{{Chainakun} \&
  {Young}}{2015}]{Chainakun2015}
{Chainakun} P.,  {Young} A.~J.,  2015, \mn@doi [\mnras]
  {10.1093/mnras/stv1333}, \href
  {http://adsabs.harvard.edu/abs/2015MNRAS.452..333C} {452, 333}

\bibitem[\protect\citeauthoryear{{Chainakun}, {Young}  \& {Kara}}{{Chainakun}
  et~al.}{2016}]{Chainakun2016}
{Chainakun} P.,  {Young} A.~J.,   {Kara} E.,  2016, \mn@doi [\mnras]
  {10.1093/mnras/stw1105}, \href
  {http://adsabs.harvard.edu/abs/2016MNRAS.460.3076C} {460, 3076}

\bibitem[\protect\citeauthoryear{{Cunningham}}{{Cunningham}}{1975}]{Cunningham%
1975}
{Cunningham} C.~T.,  1975, \mn@doi [\apj] {10.1086/154033}, \href
  {http://adsabs.harvard.edu/abs/1975ApJ...202..788C} {202, 788}

\bibitem[\protect\citeauthoryear{{Done}, {Jin}, {Middleton}  \& {Ward}}{{Done}
  et~al.}{2013}]{Done2013}
{Done} C.,  {Jin} C.,  {Middleton} M.,   {Ward} M.,  2013, \mn@doi [\mnras]
  {10.1093/mnras/stt1138}, \href
  {http://adsabs.harvard.edu/abs/2013MNRAS.434.1955D} {434, 1955}

\bibitem[\protect\citeauthoryear{{Dovciak} et~al.,}{{Dovciak}
  et~al.}{2013}]{Dovciak2013}
{Dovciak} M.,  et~al., 2013, preprint, \href
  {http://adsabs.harvard.edu/abs/2013arXiv1306.2331D} {} (\mn@eprint {arXiv}
  {1306.2331})

\bibitem[\protect\citeauthoryear{{Emmanoulopoulos}, {Papadakis}, {Dov{\v c}iak}
   \& {McHardy}}{{Emmanoulopoulos} et~al.}{2014}]{Emmanoulopoulos2014}
{Emmanoulopoulos} D.,  {Papadakis} I.~E.,  {Dov{\v c}iak} M.,   {McHardy}
  I.~M.,  2014, \mn@doi [\mnras] {10.1093/mnras/stu249}, \href
  {http://adsabs.harvard.edu/abs/2014MNRAS.439.3931E} {439, 3931}

\bibitem[\protect\citeauthoryear{{Epitropakis} \& {Papadakis}}{{Epitropakis} \&
  {Papadakis}}{2016}]{Epitropakis2016}
{Epitropakis} A.,  {Papadakis} I.~E.,  2016, \mn@doi [\aap]
  {10.1051/0004-6361/201527665}, \href
  {http://adsabs.harvard.edu/abs/2016A%26A...591A.113E} {591, A113}

\bibitem[\protect\citeauthoryear{{Epitropakis}, {Papadakis}, {Dov{\v c}iak},
  {Pech{\'a}{\v c}ek}, {Emmanoulopoulos}, {Karas}  \& {McHardy}}{{Epitropakis}
  et~al.}{2016}]{Epitropakis2016b}
{Epitropakis} A.,  {Papadakis} I.~E.,  {Dov{\v c}iak} M.,  {Pech{\'a}{\v c}ek}
  T.,  {Emmanoulopoulos} D.,  {Karas} V.,   {McHardy} I.~M.,  2016, preprint,
  \href {http://adsabs.harvard.edu/abs/2016arXiv160702625E} {} (\mn@eprint
  {arXiv} {1607.02625})

\bibitem[\protect\citeauthoryear{{Fabian} et~al.,}{{Fabian}
  et~al.}{2009}]{Fabian2009}
{Fabian} A.~C.,  et~al., 2009, \mn@doi [\nat] {10.1038/nature08007}, \href
  {http://adsabs.harvard.edu/abs/2009Natur.459..540F} {459, 540}

\bibitem[\protect\citeauthoryear{{Fanton}, {Calvani}, {de Felice}  \&
  {Cadez}}{{Fanton} et~al.}{1997}]{Fanton1997}
{Fanton} C.,  {Calvani} M.,  {de Felice} F.,   {Cadez} A.,  1997, \pasj, \href
  {http://adsabs.harvard.edu/abs/1997PASJ...49..159F} {49, 159}

\bibitem[\protect\citeauthoryear{{Garc{\'{\i}}a} et~al.,}{{Garc{\'{\i}}a}
  et~al.}{2014}]{Garcia2014}
{Garc{\'{\i}}a} J.,  et~al., 2014, \mn@doi [\apj] {10.1088/0004-637X/782/2/76},
  \href {http://adsabs.harvard.edu/abs/2014ApJ...782...76G} {782, 76}

\bibitem[\protect\citeauthoryear{{George} \& {Fabian}}{{George} \&
  {Fabian}}{1991}]{George1991}
{George} I.~M.,  {Fabian} A.~C.,  1991, \mnras, \href
  {http://adsabs.harvard.edu/abs/1991MNRAS.249..352G} {249, 352}

\bibitem[\protect\citeauthoryear{{Houck} \& {Denicola}}{{Houck} \&
  {Denicola}}{2000}]{Houck2000}
{Houck} J.~C.,  {Denicola} L.~A.,  2000, in {Manset} N.,  {Veillet} C.,
  {Crabtree} D.,  eds,  Astronomical Society of the Pacific Conference Series
  Vol. 216, Astronomical Data Analysis Software and Systems IX. p.~591

\bibitem[\protect\citeauthoryear{Jansen et~al.,}{Jansen
  et~al.}{2001}]{Jansen2001}
Jansen F.,  et~al., 2001, \mn@doi [Astron. Astrophys.]
  {10.1051/0004-6361:20000036}, 365, L1

\bibitem[\protect\citeauthoryear{{Jin}, {Done}, {Middleton}  \& {Ward}}{{Jin}
  et~al.}{2013}]{Jin2013}
{Jin} C.,  {Done} C.,  {Middleton} M.,   {Ward} M.,  2013, \mn@doi [\mnras]
  {10.1093/mnras/stt1801}, \href
  {http://adsabs.harvard.edu/abs/2013MNRAS.436.3173J} {436, 3173}

\bibitem[\protect\citeauthoryear{{Kara}, {Fabian}, {Cackett}, {Uttley},
  {Wilkins}  \& {Zoghbi}}{{Kara} et~al.}{2013}]{Kara2013a}
{Kara} E.,  {Fabian} A.~C.,  {Cackett} E.~M.,  {Uttley} P.,  {Wilkins} D.~R.,
  {Zoghbi} A.,  2013, \mn@doi [\mnras] {10.1093/mnras/stt1055}, \href
  {http://adsabs.harvard.edu/abs/2013MNRAS.434.1129K} {434, 1129}

\bibitem[\protect\citeauthoryear{{Kara}, {Cackett}, {Fabian}, {Reynolds}  \&
  {Uttley}}{{Kara} et~al.}{2014}]{Kara2014}
{Kara} E.,  {Cackett} E.~M.,  {Fabian} A.~C.,  {Reynolds} C.,   {Uttley} P.,
  2014, \mn@doi [\mnras] {10.1093/mnrasl/slt173}, \href
  {http://adsabs.harvard.edu/abs/2014MNRAS.439L..26K} {439, L26}

\bibitem[\protect\citeauthoryear{{Kara}, {Alston}  \& {Fabian}}{{Kara}
  et~al.}{2016}]{Kara2016}
{Kara} E.,  {Alston} W.,   {Fabian} A.,  2016, \mn@doi [Astronomische
  Nachrichten] {10.1002/asna.201612332}, \href
  {http://adsabs.harvard.edu/abs/2016AN....337..473K} {337, 473}

\bibitem[\protect\citeauthoryear{{Karas}, {Vokrouhlicky}  \&
  {Polnarev}}{{Karas} et~al.}{1992}]{Karas1992}
{Karas} V.,  {Vokrouhlicky} D.,   {Polnarev} A.~G.,  1992, \mnras, \href
  {http://adsabs.harvard.edu/abs/1992MNRAS.259..569K} {259, 569}

\bibitem[\protect\citeauthoryear{{Kotov}, {Churazov}  \& {Gilfanov}}{{Kotov}
  et~al.}{2001}]{Kotov2001}
{Kotov} O.,  {Churazov} E.,   {Gilfanov} M.,  2001, \mn@doi [\mnras]
  {10.1046/j.1365-8711.2001.04769.x}, \href
  {http://adsabs.harvard.edu/abs/2001MNRAS.327..799K} {327, 799}

\bibitem[\protect\citeauthoryear{{Lyubarskii}}{{Lyubarskii}}{1997}]{Lyubarskii%
1997}
{Lyubarskii} Y.~E.,  1997, \mnras, \href
  {http://adsabs.harvard.edu/abs/1997MNRAS.292..679L} {292, 679}

\bibitem[\protect\citeauthoryear{{Marconi}, {Axon}, {Maiolino}, {Nagao},
  {Pastorini}, {Pietrini}, {Robinson}  \& {Torricelli}}{{Marconi}
  et~al.}{2008}]{Marconi2008}
{Marconi} A.,  {Axon} D.~J.,  {Maiolino} R.,  {Nagao} T.,  {Pastorini} G.,
  {Pietrini} P.,  {Robinson} A.,   {Torricelli} G.,  2008, \mn@doi [\apj]
  {10.1086/529360}, \href {http://adsabs.harvard.edu/abs/2008ApJ...678..693M}
  {678, 693}

\bibitem[\protect\citeauthoryear{{Miniutti} \& {Fabian}}{{Miniutti} \&
  {Fabian}}{2004}]{Miniutti2004}
{Miniutti} G.,  {Fabian} A.~C.,  2004, \mn@doi [\mnras]
  {10.1111/j.1365-2966.2004.07611.x}, \href
  {http://adsabs.harvard.edu/abs/2004MNRAS.349.1435M} {349, 1435}

\bibitem[\protect\citeauthoryear{{Miyamoto}, {Kimura}, {Kitamoto}, {Dotani}  \&
  {Ebisawa}}{{Miyamoto} et~al.}{1991}]{Miyamoto1991}
{Miyamoto} S.,  {Kimura} K.,  {Kitamoto} S.,  {Dotani} T.,   {Ebisawa} K.,
  1991, \mn@doi [\apj] {10.1086/170837}, \href
  {http://adsabs.harvard.edu/abs/1991ApJ...383..784M} {383, 784}

\bibitem[\protect\citeauthoryear{{Nowak}, {Vaughan}, {Wilms}, {Dove}  \&
  {Begelman}}{{Nowak} et~al.}{1999}]{Nowak1999}
{Nowak} M.~A.,  {Vaughan} B.~A.,  {Wilms} J.,  {Dove} J.~B.,   {Begelman}
  M.~C.,  1999, \mn@doi [\apj] {10.1086/306610}, \href
  {http://adsabs.harvard.edu/abs/1999ApJ...510..874N} {510, 874}

\bibitem[\protect\citeauthoryear{{Papadakis}, {Pech{\'a}{\v c}ek}, {Dov{\v
  c}iak}, {Epitropakis}, {Emmanoulopoulos}  \& {Karas}}{{Papadakis}
  et~al.}{2016}]{Papadakis2016}
{Papadakis} I.,  {Pech{\'a}{\v c}ek} T.,  {Dov{\v c}iak} M.,  {Epitropakis} A.,
   {Emmanoulopoulos} D.,   {Karas} V.,  2016, \mn@doi [\aap]
  {10.1051/0004-6361/201527246}, \href
  {http://adsabs.harvard.edu/abs/2016A%26A...588A..13P} {588, A13}

\bibitem[\protect\citeauthoryear{{Ponti}, {Papadakis}, {Bianchi}, {Guainazzi},
  {Matt}, {Uttley}  \& {Bonilla}}{{Ponti} et~al.}{2012}]{Ponti2012}
{Ponti} G.,  {Papadakis} I.,  {Bianchi} S.,  {Guainazzi} M.,  {Matt} G.,
  {Uttley} P.,   {Bonilla} N.~F.,  2012, \mn@doi [\aap]
  {10.1051/0004-6361/201118326}, \href
  {http://adsabs.harvard.edu/abs/2012A%26A...542A..83P} {542, A83}

\bibitem[\protect\citeauthoryear{{Reynolds}, {Young}, {Begelman}  \&
  {Fabian}}{{Reynolds} et~al.}{1999}]{Reynolds1999}
{Reynolds} C.~S.,  {Young} A.~J.,  {Begelman} M.~C.,   {Fabian} A.~C.,  1999,
  \mn@doi [\apj] {10.1086/306913}, \href
  {http://adsabs.harvard.edu/abs/1999ApJ...514..164R} {514, 164}

\bibitem[\protect\citeauthoryear{{Ross} \& {Fabian}}{{Ross} \&
  {Fabian}}{2005}]{Ross2005}
{Ross} R.~R.,  {Fabian} A.~C.,  2005, \mn@doi [\mnras]
  {10.1111/j.1365-2966.2005.08797.x}, \href
  {http://adsabs.harvard.edu/abs/2005MNRAS.358..211R} {358, 211}

\bibitem[\protect\citeauthoryear{{Ross}, {Fabian}  \& {Young}}{{Ross}
  et~al.}{1999}]{Ross1999}
{Ross} R.~R.,  {Fabian} A.~C.,   {Young} A.~J.,  1999, \mn@doi [\mnras]
  {10.1046/j.1365-8711.1999.02528.x}, \href
  {http://adsabs.harvard.edu/abs/1999MNRAS.306..461R} {306, 461}

\bibitem[\protect\citeauthoryear{{Ruszkowski}}{{Ruszkowski}}{2000}]{Ruszkowski%
2000}
{Ruszkowski} M.,  2000, \mn@doi [\mnras] {10.1046/j.1365-8711.2000.02898.x},
  \href {http://adsabs.harvard.edu/abs/2000MNRAS.315....1R} {315, 1}

\bibitem[\protect\citeauthoryear{{Shakura} \& {Sunyaev}}{{Shakura} \&
  {Sunyaev}}{1973}]{Shakura1973}
{Shakura} N.~I.,  {Sunyaev} R.~A.,  1973, \aap, \href
  {http://adsabs.harvard.edu/abs/1973A%26A....24..337S} {24, 337}

\bibitem[\protect\citeauthoryear{{Silva}, {Uttley}  \& {Costantini}}{{Silva}
  et~al.}{2016}]{Silva2016}
{Silva} C.,  {Uttley} P.,   {Costantini} E.,  2016, preprint, \href
  {http://adsabs.harvard.edu/abs/2016arXiv160701065S} {} (\mn@eprint {arXiv}
  {1607.01065})

\bibitem[\protect\citeauthoryear{{Uttley}, {Cackett}, {Fabian}, {Kara}  \&
  {Wilkins}}{{Uttley} et~al.}{2014}]{Uttley2014}
{Uttley} P.,  {Cackett} E.~M.,  {Fabian} A.~C.,  {Kara} E.,   {Wilkins} D.~R.,
  2014, \mn@doi [\aapr] {10.1007/s00159-014-0072-0}, \href
  {http://adsabs.harvard.edu/abs/2014A%26ARv..22...72U} {22, 72}

\bibitem[\protect\citeauthoryear{{Vaughan}, {Edelson}, {Warwick}  \&
  {Uttley}}{{Vaughan} et~al.}{2003}]{Vaughan2003}
{Vaughan} S.,  {Edelson} R.,  {Warwick} R.~S.,   {Uttley} P.,  2003, \mn@doi
  [\mnras] {10.1046/j.1365-2966.2003.07042.x}, \href
  {http://adsabs.harvard.edu/abs/2003MNRAS.345.1271V} {345, 1271}

\bibitem[\protect\citeauthoryear{{Vestergaard} \& {Peterson}}{{Vestergaard} \&
  {Peterson}}{2006}]{Vestergaard2006}
{Vestergaard} M.,  {Peterson} B.~M.,  2006, \mn@doi [\apj] {10.1086/500572},
  \href {http://adsabs.harvard.edu/abs/2006ApJ...641..689V} {641, 689}

\bibitem[\protect\citeauthoryear{{Wilkins} \& {Fabian}}{{Wilkins} \&
  {Fabian}}{2013}]{Wilkins2013}
{Wilkins} D.~R.,  {Fabian} A.~C.,  2013, \mn@doi [\mnras]
  {10.1093/mnras/sts591}, \href
  {http://adsabs.harvard.edu/abs/2013MNRAS.430..247W} {430, 247}

\bibitem[\protect\citeauthoryear{{Wilkins}, {Cackett}, {Fabian}  \&
  {Reynolds}}{{Wilkins} et~al.}{2016}]{Wilkins2016}
{Wilkins} D.~R.,  {Cackett} E.~M.,  {Fabian} A.~C.,   {Reynolds} C.~S.,  2016,
  \mn@doi [\mnras] {10.1093/mnras/stw276}, \href
  {http://adsabs.harvard.edu/abs/2016MNRAS.458..200W} {458, 200}

\bibitem[\protect\citeauthoryear{{Zoghbi}, {Uttley}  \& {Fabian}}{{Zoghbi}
  et~al.}{2011}]{Zoghbi2011}
{Zoghbi} A.,  {Uttley} P.,   {Fabian} A.~C.,  2011, \mn@doi [\mnras]
  {10.1111/j.1365-2966.2010.17883.x}, \href
  {http://adsabs.harvard.edu/abs/2011MNRAS.412...59Z} {412, 59}

\bibitem[\protect\citeauthoryear{{Zoghbi}, {Reynolds}, {Cackett}, {Miniutti},
  {Kara}  \& {Fabian}}{{Zoghbi} et~al.}{2013}]{Zoghbi2013}
{Zoghbi} A.,  {Reynolds} C.,  {Cackett} E.~M.,  {Miniutti} G.,  {Kara} E.,
  {Fabian} A.~C.,  2013, \mn@doi [\apj] {10.1088/0004-637X/767/2/121}, \href
  {http://adsabs.harvard.edu/abs/2013ApJ...767..121Z} {767, 121}

\makeatother
\end{thebibliography}

\section*{Appendix}
\renewcommand\thefigure{A\arabic{figure}} 
\setcounter{figure}{0} 

\begin{figure}
    \centering
    \includegraphics*[width=70mm]{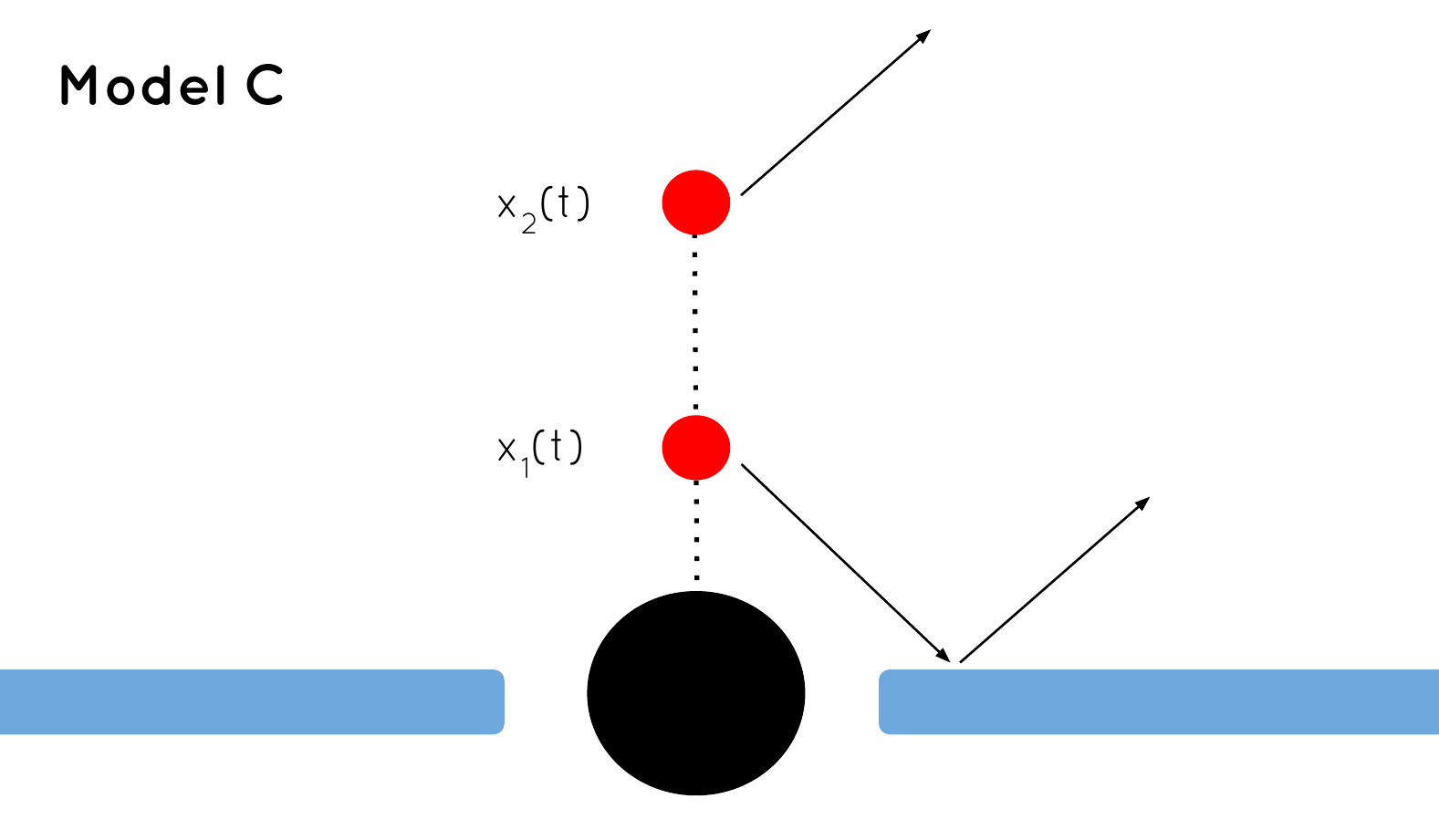}
    \vspace{-0.2cm}
    \caption{Sketch of Model C simulating the case when the observer and the disc see different parts of the X-ray sources. In this case we assume the direct photons are from only the upper source while the reflection photons are from only the lower source.} 
    \label{model_c}
\end{figure}

In this Appendix we compare the lags due to the standard and modified geometries which can be investigated by using the two-blobs model. The first scenario, referred to as Model A, is the standard one as shown in Fig.~\ref{pma} where two X-ray sources produce both direct and reflection components. The second scenario, referred to as Model B, is the geometry adopted to explain the time lags of PG~1244+026 (e.g. Fig.~\ref{geometry}) where the upper source is moving away very fast and does not produce the disc-reflection X-rays. The last scenario, referred to as Model C, demonstrates the case when the observer and the disc see different parts of the X-ray sources (i.e. different blobs, in this case). A simplistic way is to assume the disc sees only the lower source and meanwhile the observer sees only the upper source (Fig.~\ref{model_c}). In this comparison, we assume $h_{1}=5r_{\text{g}}$, $h_{2}=8r_{\text{g}}$, $i=30^{\circ}$, $A=2$, $\xi_\text{ms} = 10^4 \text{ erg cm s}^{-1}$, $p=2$, $\Gamma_{1}=2.2$, $\Gamma_{2}=2.7$, $q_{1}=0.5$, $q_{2}=1.0$, $t_{\text{max}}=1400t_{\text{g}}$ and $B=1$. Time lags are calculated in the frequency range of $(0.3-1.0)\times10^{-2} 1/t_{\text{g}}$. 

\begin{figure}
    \centering
    \includegraphics*[width=70mm]{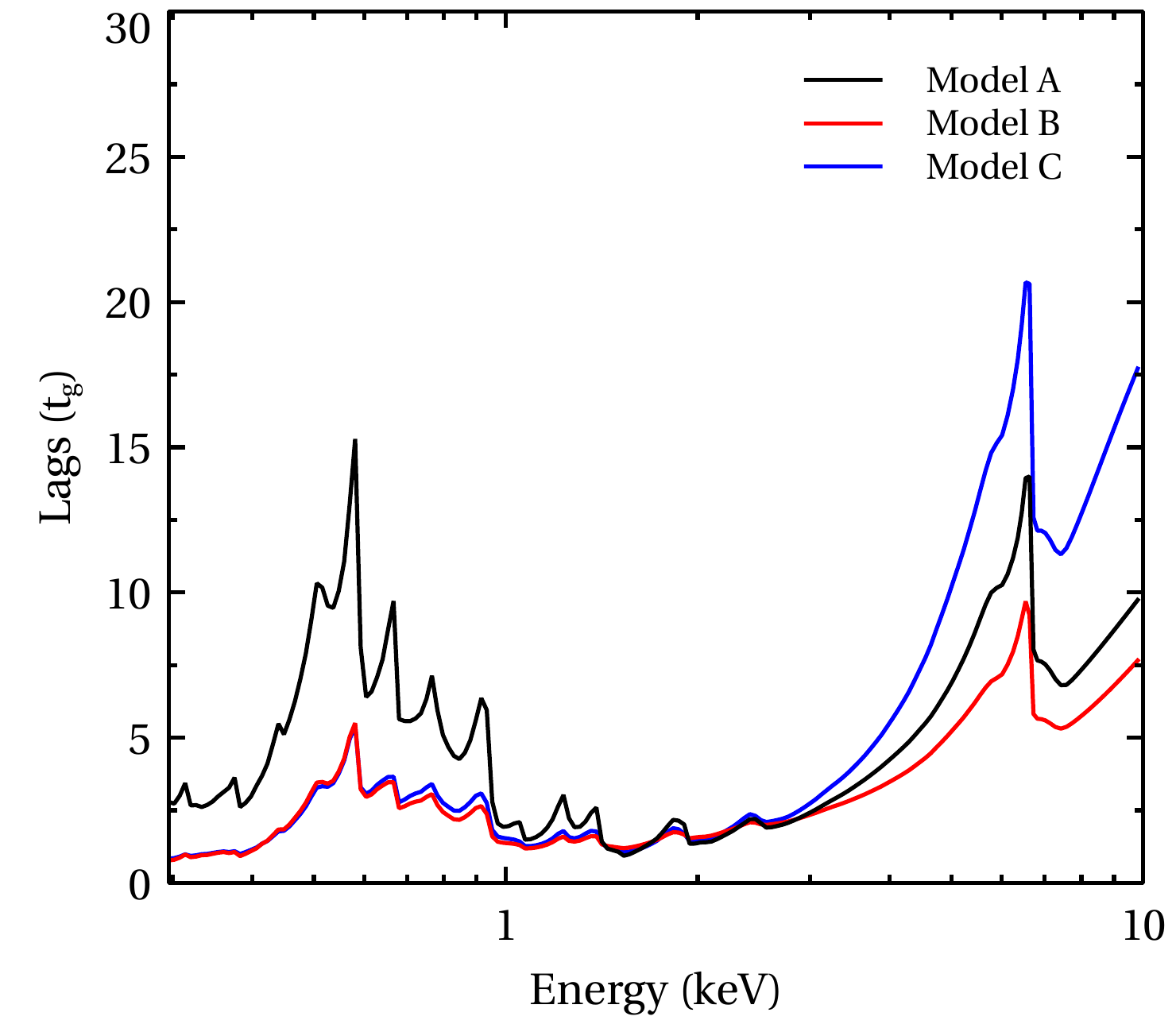}
    \vspace{-0.2cm}
    \caption{Energy-dependent reverberation lags comparing between different source-geometries when the upper source is softer than the lower source ($\Gamma_{1}=2.2$ and $\Gamma_{2}=3.0$). See Appendix for more details.} 
    \label{p10b}
\end{figure}

A comparison between the three models is shown in Fig.~\ref{p10b}. If the upper source does not produce any reflection spectrum (Model B), its soft continuum will dilute the lags between the direct and reflection components of the lower source. It is clear that the lags of Model B are those of Model A with stronger dilution in the soft excess rather than in the Fe K bands. This is because we assume the upper source emits softer X-rays than the lower source, so the soft band lags are more diluted than the hard band lags. In case of Model C, the observer sees the harder reflection and softer continuum components from the lower and upper sources, respectively. The RRF then systematically increases with energy enhancing the lags between the soft and the Fe K bands. Reversed behaviours of time lags are expected if the upper source is harder than the lower sources (i.e. dilution takes place more at the Fe K than the soft excess bands for Model B and the soft excess more enhances than the Fe K bands for Model C).  

\label{lastpage}

\end{document}